%% file: main.tex
\documentclass[letterpaper]{article} %
\usepackage{aaai25}  %
\usepackage{times}  %
\usepackage{helvet}  %
\usepackage{courier}  %
\usepackage[hyphens]{url}  %
\usepackage{graphicx} %
\usepackage{natbib}  %
\usepackage{caption} %
\usepackage{algorithm}
\usepackage{algorithmic}

\usepackage{amssymb,amsmath,amsfonts}
\usepackage[dvipsnames]{xcolor}
\usepackage{tikz}
\usepackage{booktabs}
\usepackage{subcaption}
\usepackage{makecell}
\usepackage{tabularx}
\usepackage{lipsum}
\usepackage[export]{adjustbox}
\usepackage[framemethod=TikZ]{mdframed}
\usepackage{pgfplots}
\usepackage{pgfplotstable}
\pgfplotsset{compat=1.18}
\usepgfplotslibrary{fillbetween}
\usepgfplotslibrary{external}
\usepgfplotslibrary{groupplots}
\usetikzlibrary{patterns}
\usetikzlibrary{overlay-beamer-styles}
\usetikzlibrary{arrows.meta}
\tikzexternaldisable

\newcommand{\ie}{\textit{i.e.}~}

\newcommand{\token}[1]{\texttt{<#1>}}

\newcommand{\xabs}{\mathrm{\mathbf{x}}}
\newcommand{\xref}{\mathrm{\mathbf{x}}_\text{ref}}
\newcommand{\xrelative}{\mathrm{\mathbf{x}}_\text{relative}}
\newcommand{\refelt}[2]{#1_{\text{ref}_{#2}}}
\newcommand{\Iref}{\mathrm{I}_\text{ref}}

\usepackage{newfloat}
\usepackage{listings}
\DeclareCaptionStyle{ruled}{labelfont=normalfont,labelsep=colon,strut=off} %
\floatstyle{ruled}
\newfloat{listing}{tb}{lst}{}
\floatname{listing}{Listing}
\title{Evaluating Interval-based Tokenization for Pitch Representation\\ in Symbolic Music Analysis}
\author{
    Dinh-Viet-Toan Le\textsuperscript{\rm 1}, 
    Louis Bigo\textsuperscript{\rm 2}, 
    Mikaela Keller\textsuperscript{\rm 1}
}
\affiliations{
    \textsuperscript{\rm 1}Univ. Lille, CNRS, Inria, Centrale Lille, UMR 9189 CRIStAL, F-59000 Lille\\
    \textsuperscript{\rm 2}Univ. Bordeaux, CNRS, Bordeaux INP, LaBRI, UMR 5800, F-33400 Talence\\

    dinhviettoan.le@univ-lille.fr
}

\def\StripPrefix#1>{}
\def\isOverleaf{\fi
    \def\overleafJobname{output}%
    \edef\overleafJobname{\expandafter\StripPrefix\meaning\overleafJobname}%
    \edef\job{\jobname}%
    \ifx\job\overleafJobname
}

\if\isOverleaf%
  \def\draft{1} %
\else
  \def\draft{0} %
\fi

\newcommand{\inputdraft}[2]{
\if 1\draft
    #1
\else
    \tikzexternalenable
    #2
    \tikzexternaldisable
\fi
}

\begin{document}

\maketitle

\begin{abstract}
Symbolic music analysis tasks are often performed by models originally developed for Natural Language Processing, such as Transformers.
Such models require the input data to be represented as sequences, which is achieved through a process of \emph{tokenization}.
Tokenization strategies for symbolic music often rely on absolute MIDI values to represent pitch information. 
However, music research largely promotes the benefit of higher-level representations such as melodic contour and harmonic relations for which pitch intervals turn out to be more expressive than absolute pitches.
In this work, we introduce a general framework for building interval-based tokenizations. 
By evaluating these tokenizations on three music analysis tasks, we show that such interval-based tokenizations improve model performances and facilitate their explainability.
\end{abstract}

\section{Introduction}
\label{sec:intro}

Originating from the field of Natural Language Processing (NLP), the term \emph{tokenization} initially refers to the representation of a textual content in a sequential format. The field of Music Information Retrieval has then largely adopted this term to describe the process of representing symbolic music as sequences of tokens~\cite{le2024natural}. 
This sequential representation of music enables its processing through sequential models widely adopted in the field of NLP, such as Transformers.
In contrast to text tokens, musical time and pitch are represented by metric values, raising the question of \emph{absolute} versus \emph{relative} representations.
While time encoding strategies have been studied and compared~\cite{fradet2023impact}, traditional tokenization methods mostly use \emph{absolute} pitch encoding, 
which may overlook relational aspects between notes.

Instead, music is often memorized by its melodic contour, considered as a sequence of intervals unchanged by transposition to different keys, rather than by their absolute pitches~\cite{dowling1971contour}. 
Similarly, in the context of tonal music, harmony is based on the relation between the notes constituting a chord and a tonal center more than their absolute pitches. 
Musical \emph{intervals} capture the relative distances between pitches, emphasizing relationships over fixed pitches, which aligns more closely with human musical perception. Applied to symbolic music representation, interval-based tokenization may provide a more intuitive approach to pitch encoding.

In this study, we present tokenization strategies based on intervals that can be used jointly with absolute pitch encodings. Such tokenizations are shown to improve model performances in analysis tasks.
This paper's contribution is two-fold:
\begin{itemize}
    \item We first propose a general framework to build interval-based tokenization strategies.
    \item We then show that interval-based tokenization can improve model performances on three downstream tasks. Moreover, we show that among the studied interval-based encoding strategies, optimal tokenization settings depend on the downstream task and can result in musically meaningful interpretations.
\end{itemize}

\section{Symbolic music tokenization}
\label{sec:related_works}

\input{figs/fig_intervalization.tex}

Most tokenization strategies rely on absolute pitch encodings.
Pitches are usually encoded based on MIDI numbers~\cite{huang2018music,hsiao2021compound}, as used in the REMI tokenization~\cite{huang2020pop} shown in Figure~\ref{fig:example_intervalization}. Various alternatives relying on an absolute reference have been proposed, in particular leveraging the equivalence where a pitch can be decomposed into a \token{pitch-class} and \token{octave} token~\cite{li2023pitchclass}. 
Other pitch decompositions can result from specific instrument practice, such as the DadaGP tokenization~\cite{sarmento2021dadagp} dedicated to guitar tablatures, in which pitches are encoded as pairs of string and fret values.

Some other studies have considered an encoding based on \emph{intervals} instead of absolute pitches.
In the context of monophonic songs, successive notes can be described in relation to the previous notes through intervals. Among various other viewpoints, intervals can be used to represent symbolic music in a monophonic music genre classification task~\cite{conklin2013multiple}. 
This interval-based encoding enables the discovery of ``musical words'' in monophonic music~\cite{park2024mel2word} using Byte-Pair Encoding~\cite{sennrich2016neural}, which tend to appear distinctly in specific musical styles.

Considering intervals in the more general context of polyphonic music is less straightforward because of the distinction between \emph{melodic} intervals, between successive notes, and \emph{harmonic} intervals, between simultaneous notes. To the best of our knowledge, only \citet{kermarec2022improving} have attempted a tokenization of polyphonic pieces with pitch values decribed with intervals instead. They have extended the REMI tokenization~\cite{huang2020pop} with \emph{pitch interval} tokens. In particular, they have implemented a \emph{spatial pitch-interval} tokenization strategy which distinguishes between simultaneous and consecutive notes with horizontal and vertical pitch intervals. The study is however limited in several aspects. 
On the one hand, regarding the modeling of the tokenization strategy, horizontal pitch intervals are applied to the \emph{skyline}\footnote{The \emph{skyline} of a musical content includes the highest note at every instant.} stream of notes, and all vertical pitch intervals are assumed to have a negative value. However, the skyline pitches might not always be the optimal reference notes to describe the whole musical content.
On the other hand, the experimental framework for downstream tasks is rather simple, musical content is represented as ``bag-of-tokens'' and is used as input to a logistic regression, which might not take  fully advantage of the interval-based tokenization strategy.

In this study, our aim is to extend this work by generalizing the process of encoding symbolic music using intervals instead of absolute pitches, which we call \emph{intervalization}.

\section{Intervalization}
\label{sec:intervalization}

In this section, we propose a formalization of the \emph{intervalization} process. Let $\xabs$ be a sequence of note events:
\begin{equation*}
    \xabs = \{e_1, \ldots, e_T\}
\end{equation*}
Each note event can be written as $e_k=(p_k, t_k)$ where $p_k \in \{1, \ldots, 128\}$ denotes an absolute pitch element and $t_k$ the onset time element associated to the note.

Let $\xref \subset \xabs$ be a sub-sequence of notes, chosen as \emph{reference}. $\xref$ can correspond for instance to the \emph{bottom-line} of the musical content, the \emph{skyline}, or the melody if available in the data:
\begin{align*}
    \xref & = \{\refelt{e}{1}, \ldots, \refelt{e}{\tau}\} \\
    & = \bigl\{(\refelt{p}{1}, \refelt{t}{1}), \dots, (\refelt{p}{\tau}, \refelt{t}{\tau}) \bigr\}
\end{align*}

$\xref$ is chosen to be a monophonic sequence (\ie without simultaneous events): $\refelt{t}{j} \neq \refelt{t}{j'}$ for $j, j'\in\{1, \ldots, \tau\}$. 

The choice of $\xref$ induces a partition of $\xabs$  into $\tau$ subsets of events:
\begin{equation*}
    \xabs = S_1,\ldots,S_{\tau}    
\end{equation*}
where $S_j$ is defined as the set of notes occurring between $\refelt{e}{j}$ and $\refelt{e}{j+1}$:
\begin{equation*}
   S_{j} = \left\{\refelt{e}{j}\right\} \cup 
   \left\{(p, t) \in \xabs \;\middle|\;
   \begin{aligned}
   & \refelt{t}{j} \leq t < \refelt{t}{j+1} \\
   & (p, t) \neq \refelt{e}{j}
   \end{aligned}
   \right\}
\end{equation*}

We call \emph{intervalization} $\mathrm{I}$ the process of converting an absolute pitch element into a pitch interval element. The sequence $\xabs$ is thus transformed into the sequence $\xrelative$:
\begin{equation*}
    \xrelative = \Bigl\{\mathrm{I}(e_1, \xref), \ldots, \mathrm{I}(e_T, \xref) \Bigr\}
\end{equation*}

where, for $e = (p, t) \in S_{j}$,
\begin{align*}
    \mathrm{I}(e, \xref) = 
    \begin{cases}
        \bigl(\Iref(\refelt{p}{j}, \refelt{p}{j-1}), t \bigr) & \text{if $e = \refelt{e}{j}$} 
        \\[.3em]
        \bigl(\mathrm{I}_\text{non-ref}(p, \refelt{p}{j}), t \bigr) & \text{otherwise}
    \end{cases}     
\end{align*}

where, $\Iref$ represents a method specifying the encoding method for reference pitch tokens, while $\mathrm{I}_\text{non-ref}$ represents the encoding method for non-reference tokens.

$\Iref$ can be chosen as being an encoding using absolute pitches:
$$\Iref(\refelt{p}{j}, \refelt{p}{j-1}) = \refelt{p}{j}$$
or horizontal pitch intervals~\cite{kermarec2022improving} (\ie where each pitch is encoded as a \emph{horizontal} interval with the previous pitch within the reference sequence):
$$\Iref(\refelt{p}{j}, \refelt{p}{j-1}) = \refelt{p}{j} - \refelt{p}{j-1}$$
In the latter case, the first event is dropped.  

Similarly, $\mathrm{I}_\text{non-ref}$ can be chosen as being an encoding using absolute pitches:
$$\mathrm{I}_\text{non-ref}(p, \refelt{p}{j}) = p$$
or vertical pitch intervals (\ie where each pitch is encoded as a \emph{vertical} interval in relation to the simultaneous pitch of the reference sequence):
$$\mathrm{I}_\text{non-ref}(p, \refelt{p}{j}) = p - \refelt{p}{j}$$
We give examples of these intervalization strategies in Figure~\ref{fig:example_intervalization}.

\input{figs/tab_tokenizations_description}

Although the choice of $\xref$, $\Iref$ and $\mathrm{I}_\text{non-ref}$ can be much larger as described further,
we limit this study to the six intervalization strategies listed in Table~\ref{tab:tokenizations_description} applied to the REMI tokenization strategy~\cite{huang2020pop}. 
More precisely, we altered the original REMI tokenization into a ``MIDI-score tokenization''~\cite{chou2024midibert}, defined as a tokenization strategy in which \token{Velocity} tokens are dropped.
Indeed, the datasets used for training, as presented below, are not all directly sourced from performance data, but rather from scores or generated data, where velocities do not convey real-world information. Moreover, since we do not perform generative tasks and with the downstream tasks described below being not related to musical interpretation, we assume that velocities have a limited impact on model performances.

We perform the initial tokenization strategy using the MidiTok package~\cite{fradet2021miditok}. We publicly release the datasets, source code, and pre-trained models available at \url{https://algomus.fr/code/}.

\section{Evaluation on downstream tasks}
\label{sec:evaluation}

In this section, we present an experimental framework that aims to evaluate the impact of intervalization on various MIR tasks. More precisely, our experiments are based on BERT models evaluated on three downstream tasks, namely era classification, start-of-phrase detection, and chord inversion identification.
\subsection{Downstream tasks}

We evaluate the impact of intervalization on three supervised downstream tasks associated with different datasets. 
Because we propose to evaluate the impact of intervalization when the reference is the melody, all the datasets include music with a homophonic texture, characterized by a single melody supported by an accompaniment~\cite{benward2018music}. 
In particular, we chose these datasets so that the melody is played by a single track during each piece.
A quantitative description of the datasets is given in Table~\ref{tab:datasets} (bottom).
We focus on the following three tasks:

\begin{itemize}
    \item \textbf{Start-of-phrase detection}. 
    Following the work from \citet{le2024analyzing}, the start-of-phrase detection is a sequence tagging task. The model is trained to classify each token of a sequence as being a start-of-phrase or not.
    We also follow their framework to build a synthesized dataset of folk tunes with generated piano accompaniment which includes start-of-phrase annotations. 
    In particular, we consider the ESSEN dataset~\cite{schaffrath95essen} which includes folk melodies from 47 countries 
    with phrase annotations.
    The piano accompaniment is then generated by AccoMontage~\cite{zhao2021accomontage}.
    \item \textbf{Chord inversion identification}.
    Inspired by the task of figured bass identification~\cite{ju2020automatic}, we implement a chord inversion identification task as a sequence tagging task. The model is trained to classify each token as being part of a root position, first, second, or third inversion chord.
    We consider the When-in-Rome dataset which includes roman numeral labels~\cite{gotham2023rome} from which only the chord inversion characteristic is extracted. 
    From this dataset, we only kept Bach chorales, from which we assume the melody to be the soprano voice. 
    While the dataset does include much more data than chorales, other instrumentations, such as piano solo or orchestral pieces, do not clearly involve a melodic line or a single instrument playing the melody throughout the whole piece, which can be used as a reference in our evaluation.
    \item \textbf{Era classification}. 
    This task is a binary classification task of full sequences.
    We consider the OpenScore Lieder dataset~\cite{gotham2022lieder}, which includes voice and piano pieces by composers from 1730 to 1949.
    In particular, we characterize each composer by an average year derived from their birth and death year. 
    We selected the discriminative year for this binary classification task to ensure a balanced dataset: the model is trained to classify a piece as bieng composed either before or after 1865.
\end{itemize}

\input{figs/tab_datasets}

\subsection{Model \& Pre-training}

\input{figs/fig_results_max_compare_vs_pretrain.tex}

For performing these tasks, we chose to implement a Transformer encoder-only model on which a classification layer is plugged. 
Following MidiBERT-Piano~\cite{chou2024midibert}, this last layer is a self-attention layer with a linear layer for the full sequence classification task and a pair of linear layers for the two tagging tasks. For each downstream task, we evaluate each intervalization strategy on both an end-to-end and a pre-trained model. 

We consider three datasets for pre-training for which quantitative descriptions are given in Table~\ref{tab:datasets} (top):
\begin{itemize}
    \item POP909~\cite{wang2020pop909} includes Chinese pop songs with tracks annotated as ``melody'', ``lead'' (which we also consider as melody) and ``piano''.
    \item MTC-Piano~\cite{le2024analyzing} is a dataset of Dutch folk melodies extracted from the Meertens Tune Collections~\cite{van2014meertens} with piano arrangements generated by the AccoMontage model~\cite{zhao2021accomontage}.
    \item The OpenScore String quartets collection~\cite{gotham2023openscore} features string quartets from European composers spanning the classical to late-romantic periods. The melody is approximated as the part played by the first violin.
\end{itemize}

Using the union of these corpora as a pre-training dataset, we train a Transformer encoder-only model on an unsupervised masked language model pre-training task~\cite{devlin2018bert}.
We pre-train seven models, one for each tokenization strategy (Table~\ref{tab:tokenizations_description}).
The implementation of the model is based on the MidiBERT-Piano model~\cite{chou2024midibert}.
However, while the latter consists of 12 layers with 12 heads each, we use a smaller model with 3 layers and 8 heads per layer. 

This configuration results in a model with 14M parameters, which is eight times lighter than MidiBERT-Piano. %
Thus, using our training hyperparameters, two models can fit into a single 12GB Tesla P100 GPU. The models are pre-trained until an early stopping on the validation accuracy of 10 epochs, resulting in approximately one week of pre-training for all the seven models on our hardware.

The pre-trained model weights are not frozen during the fine-tuning process.
We stop the training after an early stopping of 3 epochs.

\section{Results}
\label{sec:results}

We evaluate models on the above downstream tasks, with various settings regarding intervalization strategies presented in Table~\ref{tab:tokenizations_description}, namely, an absolute tokenization and two intervalized models with 3 references each.
For each downstream task, we evaluate each tokenization strategy on both an end-to-end and a pre-trained model. 
Each model is trained and evaluated on three seeds of the dataset splits.
Therefore, in total, 42 trainings have been performed on each downstream tasks (12 for each type of reference, 6 trainings without intervalization).

\paragraph{Impact of intervalization.}

Intervalization improves the model performance for all the tasks, both for pre-trained and end-to-end models (Figure~\ref{fig:results_max_compare_vs_pretrain}). 
However, such improvements range from a marginal 1.2\% performance increase in the case of a pre-trained model on era classification to a significant 6\% increase with an end-to-end model trained on start-of-phrase detection.
Moreover, our results show that pre-training models systematically outperform end-to-end models. On the three tasks, pre-trained model performances are on average 1.20 times better than their end-to-end counterparts, with variations depending on the task.

\input{figs/fig_results_best_ref.tex}

\paragraph{Impact of intervalization references.}

For each task, for each split of the dataset among the three considered seeds, we compare twice, once for the end-to-end and once for the fine-tuned, the models trained without intervalization to the three models trained using intervalized tokenizations based on one of the three types of references, for a given setting of $\mathrm{I}_\text{non-ref}$ and $\Iref$.
We then count the number of times that the choice of a particular reference leads to the best result among these four models. 
In total, we therefore proceed to 12 comparisons per task (3 splits $\times$ (2 pre-trained models + 2 end-to-end models)). Each comparison involves 4 models (3 intervalized + 1 absolute).
The counts of the best models associated with their reference are shown in Figure~\ref{fig:results_best_ref}.

Models trained with a melodic reference achieve the best performance in 11 of 12 comparisons for the start-of-phrase detection task.
In contrast with the skyline or bottom-line references, the melody plays a musically meaningful role.
Regarding musical phrases, Arnold Schoenberg stated~\cite[p.~3]{schoenberg1999fundamentals}:
\begin{quote}
    \it
    Phrase endings may be marked by a [...] melodic relaxation through a drop in pitch, the use of smaller intervals and fewer notes;
\end{quote}

For the era classification task, the choice of the intervalization reference does not show a significant impact on the model performance.
Unlike the other tasks, which involve local token tagging, this task focuses on classifying entire sequences into more abstract classes. Such a higher-level task may explain why tokenization plays a less critical role in this context.

Finally, for the task of chord inversion identification, the bottom-line reference leads to the best models in 10 of the 12 comparisons.
In the next paragraph, we investigate potential explanations for the effectiveness of this intervalization reference compared to the others.

\paragraph{Musical attributes reflected by the intervalization.}

We focus on the task of chord inversion identification and we analyze how a tokenizer with a bottom-line reference classifies each token.
We study the frequency of vertical pitch tokens classified by the model as root position, first, second, or third inversion (Figure~\ref{fig:inversion_interpretability}). 
This shows that particular sets of vertical pitch intervals are more prominent within specific inversions. 
In particular, these more common interval values match the musical definitions of chord inversions. For example, a major (resp. minor) third in combination with a fifth in relation to the bass note define a root position major chord (resp. minor chord) (Figure~\ref{fig:freq_inversion0}).
The presence of occurrences outside these musical definitions of major/minor/dominant chords can reflect the presence of chord extensions or note embellishments such as passing notes.
Moreover, analyzing the proportions of false positives among the predictions can explain some of the model's errors. For example, for first inversions (Figure~\ref{fig:freq_inversion1}), the largest numbers of false positives occur with thirds and sixths, which are intervals that also compose chords in root position and second inversion respectively.

Going further, various four-part writing principles~\cite{peters2016fundamentals,benward2018music} can be inferred from these frequency distributions.
For example, Figures~\ref{fig:freq_inversion0} and \ref{fig:freq_inversion1} show that third intervals (3m and 3M) occur more often with an additional octave than within the same octave. This aligns with voice spacing rules, which typically recommend that bass and tenor voices are not too close. 
Similarly, Figure~\ref{fig:freq_inversion3} shows that there is no octave doubling (P8) of the bass in the case of a third inversion. That also aligns with the fact that the seventh of a dominant seventh chord should not be doubled in four-part harmony.

\input{figs/fig_inversion_interpretability.tex}

\section{Conclusion \& future directions}

In this work, we present a framework for building interval-based tokenization strategies.
These tokenizations rely on the choice of a reference within the sequence of notes, an encoding of this reference, and an encoding of non-reference notes.
We study the case of tokenizations with an absolute reference and vertical pitch intervals for non-reference notes, as well as tokenizations based on horizontal and vertical pitch intervals. These references are chosen as the melody, skyline and bottom-line notes.

By evaluating these tokenization strategies on three downstream analysis tasks, we show that intervalization improves the models performance compared to an absolute encoding.
Moreover, the choice of a specific intervalization reference has an impact on specific tasks and can provide musically meaningful interpretations.

\subsection{Towards further interval-based tokenization}

In this work, we only studied six intervalization strategies. However, several other tokenizations based on intervals can be constructed from the formal description presented above. 

We have restricted intervalization strategies to only three types of $\xref$.
In particular, we have assumed $\xref \subset \xabs$. 
By releasing this, $\xref$ can be chosen as a musically meaningful reference such as a reference sequence composed only of the tonal center of a piece, which may help in tasks such as harmonic analysis.

In addition, we have implemented only two types of interval encodings, namely horizontal and vertical pitch intervals. 
However, similarly to pitches which can be encoded as \token{pitch-class} and \token{octave}~\cite{li2023pitchclass}, an interval can also be considered as \token{octave-interval} and \token{interval-class} as presented in Figure~\ref{fig:example_intervalization_future}. Such an intervalization strategy may allow to directly disentangle octave relations between the notes and interval classes.

\input{figs/fig_intervalization_future.tex}

Moreover, while interval-based tokenizations can be used for analysis tasks, they cannot all be used for generation purposes, in particular for tokenizations where $\Iref$ are horizontal pitch intervals because the generated sequence will not refer to a unique musical sentence. 
Consequently, tokenizations can be considered where absolute pitches are periodically indicated (for example, at the beginning of each bar), and with intervals representing horizontal/melodic and vertical/harmonic relationships relative to this periodic absolute reference.

Finally, we have only considered tokenizations based on REMI. However, the intervalization process can be applied to any tokenization strategy in which pitches are encoded as absolute values. Time-related tokens (\ie score-based time encoding using \token{Bar} and \token{Position} as REMI, or performance-based time encoding in a MIDI-like tokenization using \token{Time-Shift}~\cite{oore2018time}) are not affected by pitch encodings. Therefore, a study comparing both time and pitch representations may show possible combinations of encodings resulting in a better modeling of symbolic music.

\section*{Ethical statement}

With the choice of the presented datasets, our work inherently exhibits a bias towards a Western tonal style of music. Moreover, merging several types of genres (folk, classical, pop), as well as the construction of a dataset of folk tunes with pop accompaniment can be questioned. 
However, restraining the scope of this study can face issues, particularly due to the lack of large-scale annotated symbolic data required to train the presented models.

Similarly to several deep learning studies, our work may have an energy consumption impact due to the needed computation power for model development, training, and evaluation. This impact is, in particular, important for the pre-training phase of our models.
Although we did not precisely monitor any hardware power consumption during this study, an approximation\footnote{\url{https://mlco2.github.io/impact}} of a one-week long pre-training on our hardware reaches a consumption of around 2 $\text{kgCO}_{2}$ eq.
However, we try to limit this impact by focusing on small architectures, resulting in fewer parameters in the model as well as  shorter training times.

\bibliography{references}

\end{document}

%% file: figs/fig_intervalization.tex
\begin{figure}[t]
    \centering
    \includegraphics[width=0.5\linewidth]{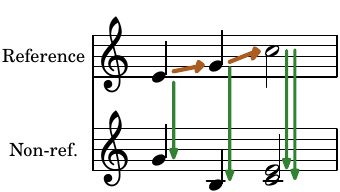}

    \vspace{1em}

    \resizebox{\linewidth}{!}{
        \input{figs/tikz_intervalization_strategies_remi.tex}
    }

    \vspace{.5em}

    {\scriptsize \textbf{Pitch intervals}: V.P.I / H.P.I (Vertical / Horizontal Pitch Interval)}

    \vspace{.5em}

    \resizebox{\linewidth}{!}{
        \input{figs/tikz_intervalization_strategies_interval_pitch.tex}
    }

    \vspace{.25em}

    \caption{
        Representations of the sheet music based on absolute and different variants of intervalization of the REMI tokenization. 
        (Abs.: \textit{Absolute pitch encoding})
    }
    \label{fig:example_intervalization}
\end{figure}

%% file: figs/tikz_intervalization_strategies_remi.tex
\tikzset{every picture/.style={line width=0.75pt}} %

\begin{tikzpicture}[x=0.75pt,y=0.75pt,yscale=-1,xscale=1]

\draw  [fill={rgb, 255:red, 120; green, 175; blue, 227 }  ,fill opacity=1 ]  (152.56,10.07) .. controls (152.56,7.31) and (154.79,5.07) .. (157.56,5.07) -- (177.56,5.07) .. controls (180.32,5.07) and (182.56,7.31) .. (182.56,10.07) -- (182.56,74.07) .. controls (182.56,76.83) and (180.32,79.07) .. (177.56,79.07) -- (157.56,79.07) .. controls (154.79,79.07) and (152.56,76.83) .. (152.56,74.07) -- cycle  ;
\draw (167.56,42.07) node  [rotate=-270] [align=left] {\begin{minipage}[lt]{47.7pt}\setlength\topsep{0pt}
\begin{center}
Dur. \ 4
\end{center}

\end{minipage}};
\draw  [fill={rgb, 255:red, 120; green, 175; blue, 227 }  ,fill opacity=1 ]  (302.42,10.07) .. controls (302.42,7.31) and (304.66,5.07) .. (307.42,5.07) -- (327.42,5.07) .. controls (330.18,5.07) and (332.42,7.31) .. (332.42,10.07) -- (332.42,74.07) .. controls (332.42,76.83) and (330.18,79.07) .. (327.42,79.07) -- (307.42,79.07) .. controls (304.66,79.07) and (302.42,76.83) .. (302.42,74.07) -- cycle  ;
\draw (317.42,42.07) node  [rotate=-270] [align=left] {\begin{minipage}[lt]{47.7pt}\setlength\topsep{0pt}
\begin{center}
Dur. \ 4
\end{center}

\end{minipage}};
\draw  [fill={rgb, 255:red, 120; green, 175; blue, 227 }  ,fill opacity=1 ]  (362.56,10.07) .. controls (362.56,7.31) and (364.8,5.07) .. (367.56,5.07) -- (387.56,5.07) .. controls (390.32,5.07) and (392.56,7.31) .. (392.56,10.07) -- (392.56,74.07) .. controls (392.56,76.83) and (390.32,79.07) .. (387.56,79.07) -- (367.56,79.07) .. controls (364.8,79.07) and (362.56,76.83) .. (362.56,74.07) -- cycle  ;
\draw (377.56,42.07) node  [rotate=-270] [align=left] {\begin{minipage}[lt]{47.7pt}\setlength\topsep{0pt}
\begin{center}
Dur. \ 4
\end{center}

\end{minipage}};
\draw  [fill={rgb, 255:red, 120; green, 175; blue, 227 }  ,fill opacity=1 ]  (452.44,9.78) .. controls (452.44,7.02) and (454.68,4.78) .. (457.44,4.78) -- (477.44,4.78) .. controls (480.21,4.78) and (482.44,7.02) .. (482.44,9.78) -- (482.44,73.78) .. controls (482.44,76.54) and (480.21,78.78) .. (477.44,78.78) -- (457.44,78.78) .. controls (454.68,78.78) and (452.44,76.54) .. (452.44,73.78) -- cycle  ;
\draw (467.44,41.78) node  [rotate=-270] [align=left] {\begin{minipage}[lt]{47.7pt}\setlength\topsep{0pt}
\begin{center}
Dur. \ 4
\end{center}

\end{minipage}};
\draw  [fill={rgb, 255:red, 222; green, 110; blue, 110 }  ,fill opacity=1 ]  (122.55,10.07) .. controls (122.55,7.31) and (124.79,5.07) .. (127.55,5.07) -- (147.55,5.07) .. controls (150.31,5.07) and (152.55,7.31) .. (152.55,10.07) -- (152.55,74.07) .. controls (152.55,76.83) and (150.31,79.07) .. (147.55,79.07) -- (127.55,79.07) .. controls (124.79,79.07) and (122.55,76.83) .. (122.55,74.07) -- cycle  ;
\draw (137.55,42.07) node  [rotate=-270] [align=left] {\begin{minipage}[lt]{47.7pt}\setlength\topsep{0pt}
\begin{center}
Pitch \ 64
\end{center}

\end{minipage}};
\draw  [fill={rgb, 255:red, 222; green, 110; blue, 110 }  ,fill opacity=1 ]  (272.55,10.07) .. controls (272.55,7.31) and (274.79,5.07) .. (277.55,5.07) -- (297.55,5.07) .. controls (300.31,5.07) and (302.55,7.31) .. (302.55,10.07) -- (302.55,74.07) .. controls (302.55,76.83) and (300.31,79.07) .. (297.55,79.07) -- (277.55,79.07) .. controls (274.79,79.07) and (272.55,76.83) .. (272.55,74.07) -- cycle  ;
\draw (287.55,42.07) node  [rotate=-270] [align=left] {\begin{minipage}[lt]{47.7pt}\setlength\topsep{0pt}
\begin{center}
Pitch \ 67
\end{center}

\end{minipage}};
\draw  [fill={rgb, 255:red, 222; green, 110; blue, 110 }  ,fill opacity=1 ]  (332.69,10.07) .. controls (332.69,7.31) and (334.93,5.07) .. (337.69,5.07) -- (357.69,5.07) .. controls (360.45,5.07) and (362.69,7.31) .. (362.69,10.07) -- (362.69,74.07) .. controls (362.69,76.83) and (360.45,79.07) .. (357.69,79.07) -- (337.69,79.07) .. controls (334.93,79.07) and (332.69,76.83) .. (332.69,74.07) -- cycle  ;
\draw (347.69,42.07) node  [rotate=-270] [align=left] {\begin{minipage}[lt]{47.7pt}\setlength\topsep{0pt}
\begin{center}
Pitch \ 59
\end{center}

\end{minipage}};
\draw  [fill={rgb, 255:red, 222; green, 110; blue, 110 }  ,fill opacity=1 ]  (422.64,10.07) .. controls (422.64,7.31) and (424.88,5.07) .. (427.64,5.07) -- (447.64,5.07) .. controls (450.4,5.07) and (452.64,7.31) .. (452.64,10.07) -- (452.64,74.07) .. controls (452.64,76.83) and (450.4,79.07) .. (447.64,79.07) -- (427.64,79.07) .. controls (424.88,79.07) and (422.64,76.83) .. (422.64,74.07) -- cycle  ;
\draw (437.64,42.07) node  [rotate=-270] [align=left] {\begin{minipage}[lt]{47.7pt}\setlength\topsep{0pt}
\begin{center}
Pitch \ 72
\end{center}

\end{minipage}};
\draw  [fill={rgb, 255:red, 159; green, 105; blue, 183 }  ,fill opacity=1 ]  (92.33,10.07) .. controls (92.33,7.31) and (94.57,5.07) .. (97.33,5.07) -- (117.33,5.07) .. controls (120.09,5.07) and (122.33,7.31) .. (122.33,10.07) -- (122.33,74.07) .. controls (122.33,76.83) and (120.09,79.07) .. (117.33,79.07) -- (97.33,79.07) .. controls (94.57,79.07) and (92.33,76.83) .. (92.33,74.07) -- cycle  ;
\draw (107.33,42.07) node  [rotate=-270] [align=left] {\begin{minipage}[lt]{47.7pt}\setlength\topsep{0pt}
\begin{center}
Pos. \ 0
\end{center}

\end{minipage}};
\draw  [fill={rgb, 255:red, 135; green, 105; blue, 47 }  ,fill opacity=1 ]  (52.83,10.07) .. controls (52.83,7.31) and (55.07,5.07) .. (57.83,5.07) -- (77.83,5.07) .. controls (80.59,5.07) and (82.83,7.31) .. (82.83,10.07) -- (82.83,74.07) .. controls (82.83,76.83) and (80.59,79.07) .. (77.83,79.07) -- (57.83,79.07) .. controls (55.07,79.07) and (52.83,76.83) .. (52.83,74.07) -- cycle  ;
\draw (67.83,42.07) node  [rotate=-270] [align=left] {\begin{minipage}[lt]{47.7pt}\setlength\topsep{0pt}
\begin{center}
Bar
\end{center}

\end{minipage}};
\draw  [fill={rgb, 255:red, 159; green, 105; blue, 183 }  ,fill opacity=1 ]  (242.62,10.07) .. controls (242.62,7.31) and (244.86,5.07) .. (247.62,5.07) -- (267.62,5.07) .. controls (270.38,5.07) and (272.62,7.31) .. (272.62,10.07) -- (272.62,74.07) .. controls (272.62,76.83) and (270.38,79.07) .. (267.62,79.07) -- (247.62,79.07) .. controls (244.86,79.07) and (242.62,76.83) .. (242.62,74.07) -- cycle  ;
\draw (257.62,42.07) node  [rotate=-270] [align=left] {\begin{minipage}[lt]{47.7pt}\setlength\topsep{0pt}
\begin{center}
Pos. \ 4
\end{center}

\end{minipage}};
\draw  [fill={rgb, 255:red, 159; green, 105; blue, 183 }  ,fill opacity=1 ]  (392.62,10.07) .. controls (392.62,7.31) and (394.86,5.07) .. (397.62,5.07) -- (417.62,5.07) .. controls (420.38,5.07) and (422.62,7.31) .. (422.62,10.07) -- (422.62,74.07) .. controls (422.62,76.83) and (420.38,79.07) .. (417.62,79.07) -- (397.62,79.07) .. controls (394.86,79.07) and (392.62,76.83) .. (392.62,74.07) -- cycle  ;
\draw (407.62,42.07) node  [rotate=-270] [align=left] {\begin{minipage}[lt]{47.7pt}\setlength\topsep{0pt}
\begin{center}
Pos. \ 8
\end{center}

\end{minipage}};
\draw (6.5,64.3) node [anchor=north west][inner sep=0.75pt]  [rotate=-270] [align=left] {\begin{minipage}[lt]{28.22pt}\setlength\topsep{0pt}
\begin{center}
\textbf{REMI}
\end{center}

\end{minipage}};
\draw  [fill={rgb, 255:red, 120; green, 175; blue, 227 }  ,fill opacity=1 ]  (212.56,10.07) .. controls (212.56,7.31) and (214.79,5.07) .. (217.56,5.07) -- (237.56,5.07) .. controls (240.32,5.07) and (242.56,7.31) .. (242.56,10.07) -- (242.56,74.07) .. controls (242.56,76.83) and (240.32,79.07) .. (237.56,79.07) -- (217.56,79.07) .. controls (214.79,79.07) and (212.56,76.83) .. (212.56,74.07) -- cycle  ;
\draw (227.56,42.07) node  [rotate=-270] [align=left] {\begin{minipage}[lt]{47.7pt}\setlength\topsep{0pt}
\begin{center}
Dur. \ 4
\end{center}

\end{minipage}};
\draw  [fill={rgb, 255:red, 222; green, 110; blue, 110 }  ,fill opacity=1 ]  (182.55,10.07) .. controls (182.55,7.31) and (184.79,5.07) .. (187.55,5.07) -- (207.55,5.07) .. controls (210.31,5.07) and (212.55,7.31) .. (212.55,10.07) -- (212.55,74.07) .. controls (212.55,76.83) and (210.31,79.07) .. (207.55,79.07) -- (187.55,79.07) .. controls (184.79,79.07) and (182.55,76.83) .. (182.55,74.07) -- cycle  ;
\draw (197.55,42.07) node  [rotate=-270] [align=left] {\begin{minipage}[lt]{47.7pt}\setlength\topsep{0pt}
\begin{center}
Pitch \ 67
\end{center}

\end{minipage}};
\draw  [fill={rgb, 255:red, 120; green, 175; blue, 227 }  ,fill opacity=1 ]  (512.44,9.78) .. controls (512.44,7.02) and (514.68,4.78) .. (517.44,4.78) -- (537.44,4.78) .. controls (540.21,4.78) and (542.44,7.02) .. (542.44,9.78) -- (542.44,73.78) .. controls (542.44,76.54) and (540.21,78.78) .. (537.44,78.78) -- (517.44,78.78) .. controls (514.68,78.78) and (512.44,76.54) .. (512.44,73.78) -- cycle  ;
\draw (527.44,41.78) node  [rotate=-270] [align=left] {\begin{minipage}[lt]{47.7pt}\setlength\topsep{0pt}
\begin{center}
Dur. \ 4
\end{center}

\end{minipage}};
\draw  [fill={rgb, 255:red, 222; green, 110; blue, 110 }  ,fill opacity=1 ]  (482.64,10.07) .. controls (482.64,7.31) and (484.88,5.07) .. (487.64,5.07) -- (507.64,5.07) .. controls (510.4,5.07) and (512.64,7.31) .. (512.64,10.07) -- (512.64,74.07) .. controls (512.64,76.83) and (510.4,79.07) .. (507.64,79.07) -- (487.64,79.07) .. controls (484.88,79.07) and (482.64,76.83) .. (482.64,74.07) -- cycle  ;
\draw (497.64,42.07) node  [rotate=-270] [align=left] {\begin{minipage}[lt]{47.7pt}\setlength\topsep{0pt}
\begin{center}
Pitch \ 64
\end{center}

\end{minipage}};
\draw  [fill={rgb, 255:red, 120; green, 175; blue, 227 }  ,fill opacity=1 ]  (572.44,9.78) .. controls (572.44,7.02) and (574.68,4.78) .. (577.44,4.78) -- (597.44,4.78) .. controls (600.21,4.78) and (602.44,7.02) .. (602.44,9.78) -- (602.44,73.78) .. controls (602.44,76.54) and (600.21,78.78) .. (597.44,78.78) -- (577.44,78.78) .. controls (574.68,78.78) and (572.44,76.54) .. (572.44,73.78) -- cycle  ;
\draw (587.44,41.78) node  [rotate=-270] [align=left] {\begin{minipage}[lt]{47.7pt}\setlength\topsep{0pt}
\begin{center}
Dur. \ 4
\end{center}

\end{minipage}};
\draw  [fill={rgb, 255:red, 222; green, 110; blue, 110 }  ,fill opacity=1 ]  (542.64,10.07) .. controls (542.64,7.31) and (544.88,5.07) .. (547.64,5.07) -- (567.64,5.07) .. controls (570.4,5.07) and (572.64,7.31) .. (572.64,10.07) -- (572.64,74.07) .. controls (572.64,76.83) and (570.4,79.07) .. (567.64,79.07) -- (547.64,79.07) .. controls (544.88,79.07) and (542.64,76.83) .. (542.64,74.07) -- cycle  ;
\draw (557.64,42.07) node  [rotate=-270] [align=left] {\begin{minipage}[lt]{47.7pt}\setlength\topsep{0pt}
\begin{center}
Pitch \ 60
\end{center}

\end{minipage}};

\end{tikzpicture}

%% file: figs/tikz_intervalization_strategies_interval_pitch.tex
\tikzset{every picture/.style={line width=0.75pt}} %

\begin{tikzpicture}[x=0.75pt,y=0.75pt,yscale=-1,xscale=1]

\draw  [fill={rgb, 255:red, 120; green, 175; blue, 227 }  ,fill opacity=1 ]  (149.06,7.77) .. controls (149.06,5.01) and (151.29,2.77) .. (154.06,2.77) -- (174.06,2.77) .. controls (176.82,2.77) and (179.06,5.01) .. (179.06,7.77) -- (179.06,71.77) .. controls (179.06,74.53) and (176.82,76.77) .. (174.06,76.77) -- (154.06,76.77) .. controls (151.29,76.77) and (149.06,74.53) .. (149.06,71.77) -- cycle  ;
\draw (164.06,39.77) node  [rotate=-270] [align=left] {\begin{minipage}[lt]{47.7pt}\setlength\topsep{0pt}
\begin{center}
Dur. \ 4
\end{center}

\end{minipage}};
\draw  [fill={rgb, 255:red, 120; green, 175; blue, 227 }  ,fill opacity=1 ]  (298.92,7.77) .. controls (298.92,5.01) and (301.16,2.77) .. (303.92,2.77) -- (323.92,2.77) .. controls (326.68,2.77) and (328.92,5.01) .. (328.92,7.77) -- (328.92,71.77) .. controls (328.92,74.53) and (326.68,76.77) .. (323.92,76.77) -- (303.92,76.77) .. controls (301.16,76.77) and (298.92,74.53) .. (298.92,71.77) -- cycle  ;
\draw (313.92,39.77) node  [rotate=-270] [align=left] {\begin{minipage}[lt]{47.7pt}\setlength\topsep{0pt}
\begin{center}
Dur. \ 4
\end{center}

\end{minipage}};
\draw  [fill={rgb, 255:red, 120; green, 175; blue, 227 }  ,fill opacity=1 ]  (359.06,7.77) .. controls (359.06,5.01) and (361.3,2.77) .. (364.06,2.77) -- (384.06,2.77) .. controls (386.82,2.77) and (389.06,5.01) .. (389.06,7.77) -- (389.06,71.77) .. controls (389.06,74.53) and (386.82,76.77) .. (384.06,76.77) -- (364.06,76.77) .. controls (361.3,76.77) and (359.06,74.53) .. (359.06,71.77) -- cycle  ;
\draw (374.06,39.77) node  [rotate=-270] [align=left] {\begin{minipage}[lt]{47.7pt}\setlength\topsep{0pt}
\begin{center}
Dur. \ 4
\end{center}

\end{minipage}};
\draw  [fill={rgb, 255:red, 120; green, 175; blue, 227 }  ,fill opacity=1 ]  (448.94,7.48) .. controls (448.94,4.72) and (451.18,2.48) .. (453.94,2.48) -- (473.94,2.48) .. controls (476.71,2.48) and (478.94,4.72) .. (478.94,7.48) -- (478.94,71.48) .. controls (478.94,74.24) and (476.71,76.48) .. (473.94,76.48) -- (453.94,76.48) .. controls (451.18,76.48) and (448.94,74.24) .. (448.94,71.48) -- cycle  ;
\draw (463.94,39.48) node  [rotate=-270] [align=left] {\begin{minipage}[lt]{47.7pt}\setlength\topsep{0pt}
\begin{center}
Dur. \ 4
\end{center}

\end{minipage}};
\draw  [fill={rgb, 255:red, 222; green, 110; blue, 110 }  ,fill opacity=1 ]  (119.05,7.77) .. controls (119.05,5.01) and (121.29,2.77) .. (124.05,2.77) -- (144.05,2.77) .. controls (146.81,2.77) and (149.05,5.01) .. (149.05,7.77) -- (149.05,71.77) .. controls (149.05,74.53) and (146.81,76.77) .. (144.05,76.77) -- (124.05,76.77) .. controls (121.29,76.77) and (119.05,74.53) .. (119.05,71.77) -- cycle  ;
\draw (134.05,39.77) node  [rotate=-270] [align=left] {\begin{minipage}[lt]{47.7pt}\setlength\topsep{0pt}
\begin{center}
Pitch \ 64
\end{center}

\end{minipage}};
\draw  [fill={rgb, 255:red, 222; green, 110; blue, 110 }  ,fill opacity=1 ]  (269.05,7.77) .. controls (269.05,5.01) and (271.29,2.77) .. (274.05,2.77) -- (294.05,2.77) .. controls (296.81,2.77) and (299.05,5.01) .. (299.05,7.77) -- (299.05,71.77) .. controls (299.05,74.53) and (296.81,76.77) .. (294.05,76.77) -- (274.05,76.77) .. controls (271.29,76.77) and (269.05,74.53) .. (269.05,71.77) -- cycle  ;
\draw (284.05,39.77) node  [rotate=-270] [align=left] {\begin{minipage}[lt]{47.7pt}\setlength\topsep{0pt}
\begin{center}
Pitch \ 67
\end{center}

\end{minipage}};
\draw  [fill={rgb, 255:red, 222; green, 110; blue, 110 }  ,fill opacity=1 ]  (419.14,7.77) .. controls (419.14,5.01) and (421.38,2.77) .. (424.14,2.77) -- (444.14,2.77) .. controls (446.9,2.77) and (449.14,5.01) .. (449.14,7.77) -- (449.14,71.77) .. controls (449.14,74.53) and (446.9,76.77) .. (444.14,76.77) -- (424.14,76.77) .. controls (421.38,76.77) and (419.14,74.53) .. (419.14,71.77) -- cycle  ;
\draw (434.14,39.77) node  [rotate=-270] [align=left] {\begin{minipage}[lt]{47.7pt}\setlength\topsep{0pt}
\begin{center}
Pitch \ 72
\end{center}

\end{minipage}};
\draw  [fill={rgb, 255:red, 159; green, 105; blue, 183 }  ,fill opacity=1 ]  (88.83,7.77) .. controls (88.83,5.01) and (91.07,2.77) .. (93.83,2.77) -- (113.83,2.77) .. controls (116.59,2.77) and (118.83,5.01) .. (118.83,7.77) -- (118.83,71.77) .. controls (118.83,74.53) and (116.59,76.77) .. (113.83,76.77) -- (93.83,76.77) .. controls (91.07,76.77) and (88.83,74.53) .. (88.83,71.77) -- cycle  ;
\draw (103.83,39.77) node  [rotate=-270] [align=left] {\begin{minipage}[lt]{47.7pt}\setlength\topsep{0pt}
\begin{center}
Pos. \ 0
\end{center}

\end{minipage}};
\draw  [fill={rgb, 255:red, 135; green, 105; blue, 47 }  ,fill opacity=1 ]  (49.33,7.77) .. controls (49.33,5.01) and (51.57,2.77) .. (54.33,2.77) -- (74.33,2.77) .. controls (77.09,2.77) and (79.33,5.01) .. (79.33,7.77) -- (79.33,71.77) .. controls (79.33,74.53) and (77.09,76.77) .. (74.33,76.77) -- (54.33,76.77) .. controls (51.57,76.77) and (49.33,74.53) .. (49.33,71.77) -- cycle  ;
\draw (64.33,39.77) node  [rotate=-270] [align=left] {\begin{minipage}[lt]{47.7pt}\setlength\topsep{0pt}
\begin{center}
Bar
\end{center}

\end{minipage}};
\draw  [fill={rgb, 255:red, 159; green, 105; blue, 183 }  ,fill opacity=1 ]  (239.12,7.77) .. controls (239.12,5.01) and (241.36,2.77) .. (244.12,2.77) -- (264.12,2.77) .. controls (266.88,2.77) and (269.12,5.01) .. (269.12,7.77) -- (269.12,71.77) .. controls (269.12,74.53) and (266.88,76.77) .. (264.12,76.77) -- (244.12,76.77) .. controls (241.36,76.77) and (239.12,74.53) .. (239.12,71.77) -- cycle  ;
\draw (254.12,39.77) node  [rotate=-270] [align=left] {\begin{minipage}[lt]{47.7pt}\setlength\topsep{0pt}
\begin{center}
Pos. \ 4
\end{center}

\end{minipage}};
\draw  [fill={rgb, 255:red, 159; green, 105; blue, 183 }  ,fill opacity=1 ]  (389.12,7.77) .. controls (389.12,5.01) and (391.36,2.77) .. (394.12,2.77) -- (414.12,2.77) .. controls (416.88,2.77) and (419.12,5.01) .. (419.12,7.77) -- (419.12,71.77) .. controls (419.12,74.53) and (416.88,76.77) .. (414.12,76.77) -- (394.12,76.77) .. controls (391.36,76.77) and (389.12,74.53) .. (389.12,71.77) -- cycle  ;
\draw (404.12,39.77) node  [rotate=-270] [align=left] {\begin{minipage}[lt]{47.7pt}\setlength\topsep{0pt}
\begin{center}
Pos. \ 8
\end{center}

\end{minipage}};
\draw (3,74) node [anchor=north west][inner sep=0.75pt]  [rotate=-270] [align=left] {\begin{minipage}[lt]{51.19pt}\setlength\topsep{0pt}
\begin{center}
\textbf{REMI}\\Abs. + VPI
\end{center}

\end{minipage}};
\draw  [fill={rgb, 255:red, 120; green, 175; blue, 227 }  ,fill opacity=1 ]  (209.06,7.77) .. controls (209.06,5.01) and (211.29,2.77) .. (214.06,2.77) -- (234.06,2.77) .. controls (236.82,2.77) and (239.06,5.01) .. (239.06,7.77) -- (239.06,71.77) .. controls (239.06,74.53) and (236.82,76.77) .. (234.06,76.77) -- (214.06,76.77) .. controls (211.29,76.77) and (209.06,74.53) .. (209.06,71.77) -- cycle  ;
\draw (224.06,39.77) node  [rotate=-270] [align=left] {\begin{minipage}[lt]{47.7pt}\setlength\topsep{0pt}
\begin{center}
Dur. \ 4
\end{center}

\end{minipage}};
\draw  [fill={rgb, 255:red, 120; green, 175; blue, 227 }  ,fill opacity=1 ]  (508.94,7.48) .. controls (508.94,4.72) and (511.18,2.48) .. (513.94,2.48) -- (533.94,2.48) .. controls (536.71,2.48) and (538.94,4.72) .. (538.94,7.48) -- (538.94,71.48) .. controls (538.94,74.24) and (536.71,76.48) .. (533.94,76.48) -- (513.94,76.48) .. controls (511.18,76.48) and (508.94,74.24) .. (508.94,71.48) -- cycle  ;
\draw (523.94,39.48) node  [rotate=-270] [align=left] {\begin{minipage}[lt]{47.7pt}\setlength\topsep{0pt}
\begin{center}
Dur. \ 4
\end{center}

\end{minipage}};
\draw  [fill={rgb, 255:red, 120; green, 175; blue, 227 }  ,fill opacity=1 ]  (568.94,7.48) .. controls (568.94,4.72) and (571.18,2.48) .. (573.94,2.48) -- (593.94,2.48) .. controls (596.71,2.48) and (598.94,4.72) .. (598.94,7.48) -- (598.94,71.48) .. controls (598.94,74.24) and (596.71,76.48) .. (593.94,76.48) -- (573.94,76.48) .. controls (571.18,76.48) and (568.94,74.24) .. (568.94,71.48) -- cycle  ;
\draw (583.94,39.48) node  [rotate=-270] [align=left] {\begin{minipage}[lt]{47.7pt}\setlength\topsep{0pt}
\begin{center}
Dur. \ 4
\end{center}

\end{minipage}};
\draw  [fill={rgb, 255:red, 111; green, 167; blue, 111 }  ,fill opacity=1 ]  (178.9,7.7) .. controls (178.9,4.94) and (181.14,2.7) .. (183.9,2.7) -- (203.9,2.7) .. controls (206.66,2.7) and (208.9,4.94) .. (208.9,7.7) -- (208.9,71.7) .. controls (208.9,74.46) and (206.66,76.7) .. (203.9,76.7) -- (183.9,76.7) .. controls (181.14,76.7) and (178.9,74.46) .. (178.9,71.7) -- cycle  ;
\draw (193.9,39.7) node  [rotate=-270] [align=left] {\begin{minipage}[lt]{47.78pt}\setlength\topsep{0pt}
\begin{center}
V.P.I \ +3
\end{center}

\end{minipage}};
\draw  [fill={rgb, 255:red, 111; green, 167; blue, 111 }  ,fill opacity=1 ]  (329.3,8.3) .. controls (329.3,5.54) and (331.54,3.3) .. (334.3,3.3) -- (354.3,3.3) .. controls (357.06,3.3) and (359.3,5.54) .. (359.3,8.3) -- (359.3,72.3) .. controls (359.3,75.06) and (357.06,77.3) .. (354.3,77.3) -- (334.3,77.3) .. controls (331.54,77.3) and (329.3,75.06) .. (329.3,72.3) -- cycle  ;
\draw (344.3,40.3) node  [rotate=-270] [align=left] {\begin{minipage}[lt]{47.78pt}\setlength\topsep{0pt}
\begin{center}
V.P.I \ -8
\end{center}

\end{minipage}};
\draw  [fill={rgb, 255:red, 111; green, 167; blue, 111 }  ,fill opacity=1 ]  (479.3,8.3) .. controls (479.3,5.54) and (481.54,3.3) .. (484.3,3.3) -- (504.3,3.3) .. controls (507.06,3.3) and (509.3,5.54) .. (509.3,8.3) -- (509.3,72.3) .. controls (509.3,75.06) and (507.06,77.3) .. (504.3,77.3) -- (484.3,77.3) .. controls (481.54,77.3) and (479.3,75.06) .. (479.3,72.3) -- cycle  ;
\draw (494.3,40.3) node  [rotate=-270] [align=left] {\begin{minipage}[lt]{47.78pt}\setlength\topsep{0pt}
\begin{center}
V.P.I \ -8
\end{center}

\end{minipage}};
\draw  [fill={rgb, 255:red, 111; green, 167; blue, 111 }  ,fill opacity=1 ]  (539.3,8.3) .. controls (539.3,5.54) and (541.54,3.3) .. (544.3,3.3) -- (564.3,3.3) .. controls (567.06,3.3) and (569.3,5.54) .. (569.3,8.3) -- (569.3,72.3) .. controls (569.3,75.06) and (567.06,77.3) .. (564.3,77.3) -- (544.3,77.3) .. controls (541.54,77.3) and (539.3,75.06) .. (539.3,72.3) -- cycle  ;
\draw (554.3,40.3) node  [rotate=-270] [align=left] {\begin{minipage}[lt]{47.78pt}\setlength\topsep{0pt}
\begin{center}
V.P.I \ -12
\end{center}

\end{minipage}};
\draw  [fill={rgb, 255:red, 120; green, 175; blue, 227 }  ,fill opacity=1 ]  (298.92,94.77) .. controls (298.92,92.01) and (301.16,89.77) .. (303.92,89.77) -- (323.92,89.77) .. controls (326.68,89.77) and (328.92,92.01) .. (328.92,94.77) -- (328.92,158.77) .. controls (328.92,161.53) and (326.68,163.77) .. (323.92,163.77) -- (303.92,163.77) .. controls (301.16,163.77) and (298.92,161.53) .. (298.92,158.77) -- cycle  ;
\draw (313.92,126.77) node  [rotate=-270] [align=left] {\begin{minipage}[lt]{47.7pt}\setlength\topsep{0pt}
\begin{center}
Dur. \ 4
\end{center}

\end{minipage}};
\draw  [fill={rgb, 255:red, 120; green, 175; blue, 227 }  ,fill opacity=1 ]  (359.06,94.77) .. controls (359.06,92.01) and (361.3,89.77) .. (364.06,89.77) -- (384.06,89.77) .. controls (386.82,89.77) and (389.06,92.01) .. (389.06,94.77) -- (389.06,158.77) .. controls (389.06,161.53) and (386.82,163.77) .. (384.06,163.77) -- (364.06,163.77) .. controls (361.3,163.77) and (359.06,161.53) .. (359.06,158.77) -- cycle  ;
\draw (374.06,126.77) node  [rotate=-270] [align=left] {\begin{minipage}[lt]{47.7pt}\setlength\topsep{0pt}
\begin{center}
Dur. \ 4
\end{center}

\end{minipage}};
\draw  [fill={rgb, 255:red, 120; green, 175; blue, 227 }  ,fill opacity=1 ]  (448.94,94.48) .. controls (448.94,91.72) and (451.18,89.48) .. (453.94,89.48) -- (473.94,89.48) .. controls (476.71,89.48) and (478.94,91.72) .. (478.94,94.48) -- (478.94,158.48) .. controls (478.94,161.24) and (476.71,163.48) .. (473.94,163.48) -- (453.94,163.48) .. controls (451.18,163.48) and (448.94,161.24) .. (448.94,158.48) -- cycle  ;
\draw (463.94,126.48) node  [rotate=-270] [align=left] {\begin{minipage}[lt]{47.7pt}\setlength\topsep{0pt}
\begin{center}
Dur. \ 4
\end{center}

\end{minipage}};
\draw  [fill={rgb, 255:red, 159; green, 105; blue, 183 }  ,fill opacity=1 ]  (88.83,94.77) .. controls (88.83,92.01) and (91.07,89.77) .. (93.83,89.77) -- (113.83,89.77) .. controls (116.59,89.77) and (118.83,92.01) .. (118.83,94.77) -- (118.83,158.77) .. controls (118.83,161.53) and (116.59,163.77) .. (113.83,163.77) -- (93.83,163.77) .. controls (91.07,163.77) and (88.83,161.53) .. (88.83,158.77) -- cycle  ;
\draw (103.83,126.77) node  [rotate=-270] [align=left] {\begin{minipage}[lt]{47.7pt}\setlength\topsep{0pt}
\begin{center}
Pos. \ 0
\end{center}

\end{minipage}};
\draw  [fill={rgb, 255:red, 135; green, 105; blue, 47 }  ,fill opacity=1 ]  (49.33,94.77) .. controls (49.33,92.01) and (51.57,89.77) .. (54.33,89.77) -- (74.33,89.77) .. controls (77.09,89.77) and (79.33,92.01) .. (79.33,94.77) -- (79.33,158.77) .. controls (79.33,161.53) and (77.09,163.77) .. (74.33,163.77) -- (54.33,163.77) .. controls (51.57,163.77) and (49.33,161.53) .. (49.33,158.77) -- cycle  ;
\draw (64.33,126.77) node  [rotate=-270] [align=left] {\begin{minipage}[lt]{47.7pt}\setlength\topsep{0pt}
\begin{center}
Bar
\end{center}

\end{minipage}};
\draw  [fill={rgb, 255:red, 159; green, 105; blue, 183 }  ,fill opacity=1 ]  (239.12,94.77) .. controls (239.12,92.01) and (241.36,89.77) .. (244.12,89.77) -- (264.12,89.77) .. controls (266.88,89.77) and (269.12,92.01) .. (269.12,94.77) -- (269.12,158.77) .. controls (269.12,161.53) and (266.88,163.77) .. (264.12,163.77) -- (244.12,163.77) .. controls (241.36,163.77) and (239.12,161.53) .. (239.12,158.77) -- cycle  ;
\draw (254.12,126.77) node  [rotate=-270] [align=left] {\begin{minipage}[lt]{47.7pt}\setlength\topsep{0pt}
\begin{center}
Pos. \ 4
\end{center}

\end{minipage}};
\draw  [fill={rgb, 255:red, 159; green, 105; blue, 183 }  ,fill opacity=1 ]  (389.12,94.77) .. controls (389.12,92.01) and (391.36,89.77) .. (394.12,89.77) -- (414.12,89.77) .. controls (416.88,89.77) and (419.12,92.01) .. (419.12,94.77) -- (419.12,158.77) .. controls (419.12,161.53) and (416.88,163.77) .. (414.12,163.77) -- (394.12,163.77) .. controls (391.36,163.77) and (389.12,161.53) .. (389.12,158.77) -- cycle  ;
\draw (404.12,126.77) node  [rotate=-270] [align=left] {\begin{minipage}[lt]{47.7pt}\setlength\topsep{0pt}
\begin{center}
Pos. \ 8
\end{center}

\end{minipage}};
\draw (3,161) node [anchor=north west][inner sep=0.75pt]  [rotate=-270] [align=left] {\begin{minipage}[lt]{47.78pt}\setlength\topsep{0pt}
\begin{center}
\textbf{REMI}\\HPI + VPI
\end{center}

\end{minipage}};
\draw  [fill={rgb, 255:red, 120; green, 175; blue, 227 }  ,fill opacity=1 ]  (209.06,94.77) .. controls (209.06,92.01) and (211.29,89.77) .. (214.06,89.77) -- (234.06,89.77) .. controls (236.82,89.77) and (239.06,92.01) .. (239.06,94.77) -- (239.06,158.77) .. controls (239.06,161.53) and (236.82,163.77) .. (234.06,163.77) -- (214.06,163.77) .. controls (211.29,163.77) and (209.06,161.53) .. (209.06,158.77) -- cycle  ;
\draw (224.06,126.77) node  [rotate=-270] [align=left] {\begin{minipage}[lt]{47.7pt}\setlength\topsep{0pt}
\begin{center}
Dur. \ 4
\end{center}

\end{minipage}};
\draw  [fill={rgb, 255:red, 120; green, 175; blue, 227 }  ,fill opacity=1 ]  (508.94,94.48) .. controls (508.94,91.72) and (511.18,89.48) .. (513.94,89.48) -- (533.94,89.48) .. controls (536.71,89.48) and (538.94,91.72) .. (538.94,94.48) -- (538.94,158.48) .. controls (538.94,161.24) and (536.71,163.48) .. (533.94,163.48) -- (513.94,163.48) .. controls (511.18,163.48) and (508.94,161.24) .. (508.94,158.48) -- cycle  ;
\draw (523.94,126.48) node  [rotate=-270] [align=left] {\begin{minipage}[lt]{47.7pt}\setlength\topsep{0pt}
\begin{center}
Dur. \ 4
\end{center}

\end{minipage}};
\draw  [fill={rgb, 255:red, 120; green, 175; blue, 227 }  ,fill opacity=1 ]  (568.94,94.48) .. controls (568.94,91.72) and (571.18,89.48) .. (573.94,89.48) -- (593.94,89.48) .. controls (596.71,89.48) and (598.94,91.72) .. (598.94,94.48) -- (598.94,158.48) .. controls (598.94,161.24) and (596.71,163.48) .. (593.94,163.48) -- (573.94,163.48) .. controls (571.18,163.48) and (568.94,161.24) .. (568.94,158.48) -- cycle  ;
\draw (583.94,126.48) node  [rotate=-270] [align=left] {\begin{minipage}[lt]{47.7pt}\setlength\topsep{0pt}
\begin{center}
Dur. \ 4
\end{center}

\end{minipage}};
\draw  [fill={rgb, 255:red, 111; green, 167; blue, 111 }  ,fill opacity=1 ]  (178.4,94.7) .. controls (178.4,91.94) and (180.64,89.7) .. (183.4,89.7) -- (203.4,89.7) .. controls (206.16,89.7) and (208.4,91.94) .. (208.4,94.7) -- (208.4,158.7) .. controls (208.4,161.46) and (206.16,163.7) .. (203.4,163.7) -- (183.4,163.7) .. controls (180.64,163.7) and (178.4,161.46) .. (178.4,158.7) -- cycle  ;
\draw (193.4,126.7) node  [rotate=-270] [align=left] {\begin{minipage}[lt]{47.78pt}\setlength\topsep{0pt}
\begin{center}
V.P.I \ +3
\end{center}

\end{minipage}};
\draw  [fill={rgb, 255:red, 111; green, 167; blue, 111 }  ,fill opacity=1 ]  (329.3,95.3) .. controls (329.3,92.54) and (331.54,90.3) .. (334.3,90.3) -- (354.3,90.3) .. controls (357.06,90.3) and (359.3,92.54) .. (359.3,95.3) -- (359.3,159.3) .. controls (359.3,162.06) and (357.06,164.3) .. (354.3,164.3) -- (334.3,164.3) .. controls (331.54,164.3) and (329.3,162.06) .. (329.3,159.3) -- cycle  ;
\draw (344.3,127.3) node  [rotate=-270] [align=left] {\begin{minipage}[lt]{47.78pt}\setlength\topsep{0pt}
\begin{center}
V.P.I \ -8
\end{center}

\end{minipage}};
\draw  [fill={rgb, 255:red, 111; green, 167; blue, 111 }  ,fill opacity=1 ]  (479.3,95.3) .. controls (479.3,92.54) and (481.54,90.3) .. (484.3,90.3) -- (504.3,90.3) .. controls (507.06,90.3) and (509.3,92.54) .. (509.3,95.3) -- (509.3,159.3) .. controls (509.3,162.06) and (507.06,164.3) .. (504.3,164.3) -- (484.3,164.3) .. controls (481.54,164.3) and (479.3,162.06) .. (479.3,159.3) -- cycle  ;
\draw (494.3,127.3) node  [rotate=-270] [align=left] {\begin{minipage}[lt]{47.78pt}\setlength\topsep{0pt}
\begin{center}
V.P.I \ -8
\end{center}

\end{minipage}};
\draw  [fill={rgb, 255:red, 111; green, 167; blue, 111 }  ,fill opacity=1 ]  (539.3,95.3) .. controls (539.3,92.54) and (541.54,90.3) .. (544.3,90.3) -- (564.3,90.3) .. controls (567.06,90.3) and (569.3,92.54) .. (569.3,95.3) -- (569.3,159.3) .. controls (569.3,162.06) and (567.06,164.3) .. (564.3,164.3) -- (544.3,164.3) .. controls (541.54,164.3) and (539.3,162.06) .. (539.3,159.3) -- cycle  ;
\draw (554.3,127.3) node  [rotate=-270] [align=left] {\begin{minipage}[lt]{47.78pt}\setlength\topsep{0pt}
\begin{center}
V.P.I \ -12
\end{center}

\end{minipage}};
\draw  [fill={rgb, 255:red, 230; green, 182; blue, 79 }  ,fill opacity=1 ]  (269.11,94.91) .. controls (269.11,92.15) and (271.35,89.91) .. (274.11,89.91) -- (294.11,89.91) .. controls (296.88,89.91) and (299.11,92.15) .. (299.11,94.91) -- (299.11,158.91) .. controls (299.11,161.67) and (296.88,163.91) .. (294.11,163.91) -- (274.11,163.91) .. controls (271.35,163.91) and (269.11,161.67) .. (269.11,158.91) -- cycle  ;
\draw (284.11,126.91) node  [rotate=-270] [align=left] {\begin{minipage}[lt]{47.78pt}\setlength\topsep{0pt}
\begin{center}
H.P.I \ +3
\end{center}

\end{minipage}};
\draw  [fill={rgb, 255:red, 230; green, 182; blue, 79 }  ,fill opacity=1 ]  (419.11,94.91) .. controls (419.11,92.15) and (421.35,89.91) .. (424.11,89.91) -- (444.11,89.91) .. controls (446.88,89.91) and (449.11,92.15) .. (449.11,94.91) -- (449.11,158.91) .. controls (449.11,161.67) and (446.88,163.91) .. (444.11,163.91) -- (424.11,163.91) .. controls (421.35,163.91) and (419.11,161.67) .. (419.11,158.91) -- cycle  ;
\draw (434.11,126.91) node  [rotate=-270] [align=left] {\begin{minipage}[lt]{47.78pt}\setlength\topsep{0pt}
\begin{center}
H.P.I \ +5
\end{center}

\end{minipage}};

\end{tikzpicture}

%% file: figs/tab_tokenizations_description.tex
\begin{table}[t]
    \centering
    \begin{tabular}{l|lll}
        \toprule
        \textbf{Tokenization} & $\xref$ & $\mathrm{I}_{\text{ref}}$ & $\mathrm{I}_{\text{non-ref}}$ \\
        \midrule
        REMI-absolute & -- & Abs. & Abs. \\
        \midrule
        REMI-abs.+VPI \\ 
        \quad ref-melody & Melody & Abs. & V.P.I. \\
        \quad ref-skyline & Skyline & Abs. & V.P.I. \\
        \quad ref-bottom-line & Bottom-line & Abs. & V.P.I. \\
        \midrule
        REMI-HPI+VPI \\ 
        \quad ref-melody & Melody & H.P.I. & V.P.I. \\
        \quad ref-skyline & Skyline & H.P.I. & V.P.I. \\
        \quad ref-bottom-line & Bottom-line & H.P.I. & V.P.I. \\        
        \bottomrule
    \end{tabular}
    \caption{
        Tokenizations studied in this work, including the original REMI tokenization (REMI-absolute), and interval-based tokenizations based on REMI.
        (Abs.: \textit{Absolute pitch encoding} ; V.P.I.: \textit{Vertical Pitch Interval} ; H.P.I.: \textit{Horizontal Pitch Interval})
    }
    \label{tab:tokenizations_description}
\end{table}

%% file: figs/tab_datasets.tex
\newcommand{\countTP}[2]{#1~{\small(#2)}}

\begin{table}[t]
    \centering
    \begin{tabular}{lll}
        \toprule
        \textbf{Dataset} & \textbf{Task} & \makecell{\textbf{\# tokens}\\{\small (\# pieces)}} \\
        \midrule
        POP909 & Pre-training & \countTP{12.1M}{2897} \\ %
        MTC-Piano & Pre-training & \countTP{12.4M}{18.1k} \\ %
        String quartets & Pre-training & \countTP{3.2M}{121} \\ %
        \cmidrule{3-3}
        \textit{Total} & \textit{Pre-training} & \textit{\countTP{27.8M}{21.1k}} \\ %
        \midrule
        Lieder & Era classification & \countTP{2.7M}{1356} \\ %
        ESSEN-Piano & Phrase detection & \countTP{3.4M}{6926} \\ %
        Bach chorales & Chord inv. ident. & \countTP{204k}{371} \\ %
        \bottomrule
    \end{tabular}
    \caption{
        Description of the datasets used for pre-training and downstream tasks, namely era classification, start-of-phrase detection and chord inversion identification.
        The count of tokens is given in terms of REMI-absolute tokens.
        }
    \label{tab:datasets}
\end{table}

%% file: figs/fig_results_max_compare_vs_pretrain.tex
\begin{figure}[t]
    \centering
    \inputdraft{
        \includegraphics{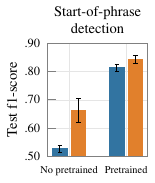}
        \hfill
        \includegraphics{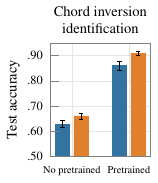}
        \hfill
        \includegraphics{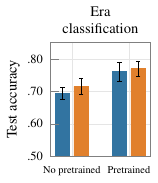}
        
        \includegraphics{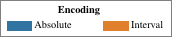}
    }{
        \input{figs/pgf_perfo.tex}
    }
    \caption{
        Performance comparison between absolute and intervalized tokenization strategies on the three downstream tasks with non pre-trained and pre-trained models.
        The intervalized model is based on the reference resulting in the best performance.
    }
    \label{fig:results_max_compare_vs_pretrain}
\end{figure}

%% file: figs/pgf_perfo.tex
\pgfplotsset{%
  every axis legend/.append style ={%
    anchor = north west,%
    at = {(0.02,0.97)}%
  },
  /pgf/number format/.cd,
  use comma,
  1000 sep = {\,},
  min exponent for 1000 sep = 4,
}

\pgfplotsset{
  discard if/.style 2 args={
    x filter/.code={
      \edef\tempa{\thisrow{#1}}
      \edef\tempb{#2}
      \ifx\tempa\tempb
      \def\pgfmathresult{inf}
      \fi
    }
  },
  discard if not/.style 2 args={
    x filter/.code={
      \edef\tempa{\thisrow{#1}}
      \edef\tempb{#2}
      \ifx\tempa\tempb
      \else
      \def\pgfmathresult{inf}
      \fi
    }
  }
}

\makeatletter \newcommand{\pgfplotsdrawaxis}{\pgfplots@draw@axis} \makeatother

\pgfplotsset{axis line on top/.style={
  axis line style=transparent,
  ticklabel style=transparent,
  tick style=transparent,
  axis on top=false,
  after end axis/.append code={
    \pgfplotsset{axis line style=opaque,
      ticklabel style=opaque,
      tick style=opaque,
      grid=none}
    \pgfplotsdrawaxis}
  }
}

\definecolor{tab-1}{HTML}{3274a1} %
\definecolor{tab-2}{HTML}{e1802c} %
\definecolor{tab-3}{HTML}{2ca02c} %
\definecolor{tab-4}{HTML}{d62728} %
\definecolor{tab-5}{HTML}{9467bd} %
\definecolor{tab-6}{HTML}{8c564b} %
\definecolor{tab-7}{HTML}{e377c2} %
\definecolor{tab-8}{HTML}{7f7f7f} %
\definecolor{tab-9}{HTML}{bcbd22} %
\definecolor{tab-10}{HTML}{17becf} %

\newcommand\tinier{\fontsize{5pt}{5pt}\selectfont}

\centering
\tikzsetnextfilename{pgf_phrase_segmentation_perfo}
\begin{tikzpicture}
  \begin{axis}[
      axis line on top,
      ybar,
      align=center,
      bar width=7pt,
      ylabel={Test f1-score},
      xlabel={},
      xtick=data,
      symbolic x coords={No pretrained, Pretrained},
      xticklabel style={align=center},
      ytick={0.50, 0.60, 00.70, 0.80, 0.90},
      yticklabels={.50, .60, .70, .80, .90},
      ymin=0.5, ymax=0.90,
      enlarge x limits=0.38,
      tick pos=left,
      cycle list name=color list,
      width=.39\linewidth,height=3.5cm,
        title style={
            yshift=-.5em
        },
      title={\scriptsize Start-of-phrase\\[-.3em] \scriptsize detection},
      axis line style = {gray},
      xtick style={draw=none},
      x tick label style={
        rotate=0,
        anchor=north,
        font=\tinier,
        yshift=.3em
      },
      y tick label style={
        rotate=0,
        font=\tiny
      },
      y label style={
        at={(axis description cs:-.23,.5)},anchor=south,
        font=\scriptsize
      },
      x label style={
        font=\scriptsize
      },
      legend style={
        nodes={scale=.6, transform shape},
        font=\ttfamily,
        legend cell align={left},
        at={(0.6,0.05)},
        anchor=south,
        legend image post style={xscale=1},
        /tikz/every even column/.append style={column sep=.2cm}
      },
      legend columns=2,
      grid=both,
      grid style={line width=.1pt, draw=gray!20},
    ]

    \addplot+ [discard if not={is_interval}{False}][
    tab-1,fill=tab-1,  
    error bars/.cd,
      y dir=both,y explicit,
      error bar style={black},
      error mark options={rotate=90, mark size=1pt}
    ]
    table[
      col sep=comma,
      x=model,
      y=test_f1_mean,
      y error=test_f1_std,
    ] {data/phrase_segmentation_perfo.csv};

    \addplot+ [discard if not={is_interval}{True}][
      tab-2,fill=tab-2,
      error bars/.cd,
      y dir=both, y explicit,
      error bar style={black},
      error mark options={rotate=90, mark size=1pt}
    ] table[
      col sep=comma,
      x=model,
      y=test_f1_mean,
      y error=test_f1_std,
      meta=is_interval,
    ] {data/phrase_segmentation_perfo.csv};

  \end{axis}
\end{tikzpicture}
\tikzsetnextfilename{pgf_chord_identification_perfo}
\begin{tikzpicture}
  \begin{axis}[
      axis line on top,
      ybar,
      align=center,
      bar width=7pt,
      ylabel={Test accuracy},
      xlabel={},
      xtick=data,
      symbolic x coords={No pretrained, Pretrained},
      xticklabel style={align=center},
      ytick={0.50, 0.60, 0.70, 0.80, 0.90},
      yticklabels={.50, .60, .70, .80, .90},
      ymin=0.5, ymax=0.95,
      enlarge x limits=0.38,
      tick pos=left,
      cycle list name=color list,
      width=.39\linewidth,height=3.5cm,
      title style={
            yshift=-.5em
        },
      title={\scriptsize Chord inversion\\[-.3em] \scriptsize identification},
      axis line style = {gray},
      x tick label style={
        rotate=0,
        anchor=north,
        font=\tinier,
        yshift=.3em
      },
      xtick style={draw=none},
      y tick label style={
        rotate=0,
        font=\tiny
      },
      y label style={
        at={(axis description cs:-.23,.5)},anchor=south,
        font=\scriptsize
      },
      x label style={
        font=\scriptsize
      },
      legend style={
        nodes={scale=.6, transform shape},
        legend cell align={center},
        at={(0.4,-0.5)},
        anchor=south,
        legend image post style={xscale=1},
        /tikz/every even column/.append style={column sep=.2cm}
      },
      legend columns=2,
      grid=both,
      grid style={line width=.1pt, draw=gray!20},
    ]

    \addplot+ [discard if not={is_interval}{False}][
    tab-1,fill=tab-1,  
    error bars/.cd,
      y dir=both,y explicit,
      error bar style={black},
      error mark options={rotate=90, mark size=1pt}
    ]
    table[
      col sep=comma,
      x=model,
      y=test_acc_mean,
      y error=test_acc_std,
    ] {data/chord_inversion_perfo.csv};

    \addplot+ [discard if not={is_interval}{True}][
      tab-2,fill=tab-2,
      error bars/.cd,
      y dir=both, y explicit,
      error bar style={black},
      error mark options={rotate=90, mark size=1pt}
    ] table[
      col sep=comma,
      x=model,
      y=test_acc_mean,
      y error=test_acc_std,
      meta=is_interval,
    ] {data/chord_inversion_perfo.csv};

  \end{axis}
\end{tikzpicture}
\tikzsetnextfilename{pgf_era_clf_perfo}
\begin{tikzpicture}
  \begin{axis}[
      axis line on top,
      ybar,
      align=center,
      bar width=7pt,
      ylabel={Test accuracy},
      xlabel={},
      xtick=data,
      symbolic x coords={No pretrained, Pretrained},
      xticklabel style={align=center},
      ytick={0.50, 0.60, 0.70, 0.80, 0.90},
      yticklabels={.50, .60, .70, .80, .90},
      ymin=0.5, ymax=0.85,
      enlarge x limits=0.38,
      tick pos=left,
      cycle list name=color list,
      width=.39\linewidth,height=3.5cm,
      title style={
            yshift=-.5em
        },
      title={\scriptsize Era\\[-.3em] \scriptsize classification},
      axis line style = {gray},
      x tick label style={
        rotate=0,
        anchor=north,
        font=\tinier,
        yshift=.3em
      },
      xtick style={draw=none},
      y tick label style={
        rotate=0,
        font=\tiny
      },
      y label style={
        at={(axis description cs:-.23,.5)},anchor=south,
        font=\scriptsize
      },
      x label style={
        font=\scriptsize
      },
      legend style={
        nodes={scale=.6, transform shape},
        font=\ttfamily,
        legend cell align={left},
        at={(0.6,0.05)},
        anchor=south,
        legend image post style={xscale=1},
        /tikz/every even column/.append style={column sep=.2cm}
      },
      legend columns=2,
      grid=both,
      grid style={line width=.1pt, draw=gray!20},
    ]

    \addplot+ [discard if not={is_interval}{False}][
    tab-1,fill=tab-1,  
    error bars/.cd,
      y dir=both,y explicit,
      error bar style={black},
      error mark options={rotate=90, mark size=1pt}
    ]
    table[
      col sep=comma,
      x=model,
      y=test_acc_mean,
      y error=test_acc_std,
    ] {data/era_clf_perfo.csv};

    \addplot+ [discard if not={is_interval}{True}][
      tab-2,fill=tab-2,
      error bars/.cd,
      y dir=both, y explicit,
      error bar style={black},
      error mark options={rotate=90, mark size=1pt}
    ] table[
      col sep=comma,
      x=model,
      y=test_acc_mean,
      y error=test_acc_std,
      meta=is_interval,
    ] {data/era_clf_perfo.csv};

  \end{axis}
\end{tikzpicture}

\tikzsetnextfilename{pgf_legend_perfo}
\begin{tikzpicture} 
    \begin{axis}[%
    hide axis,
    xmin=10,
    xmax=50,
    ymin=0,
    ymax=0.4,
    legend style={
        nodes={scale=.5, transform shape},
        legend cell align={left},
        at={(0.6,0.05)},
        anchor=south,
        legend image post style={xscale=1},
        /tikz/every even column/.append style={column sep=.5cm},
        row sep=1pt,
        legend image post style={scale=0.7},
        draw=gray
    },
    legend columns=2, 
    width=\linewidth,
    ]
    \addlegendimage{empty legend} %
    \addlegendentry{\makebox[0pt][l]{\hspace{.8cm}\textbf{Encoding}}}
    
    \addlegendimage{empty legend}
    \addlegendentry{}

    \addlegendimage{area legend,color=tab-1,fill}
    \addlegendentry{Absolute};
    \addlegendimage{area legend,color=tab-2,fill}
    \addlegendentry{Interval};
    \end{axis}
\end{tikzpicture}

%% file: figs/fig_results_best_ref.tex
\begin{figure}[t]
    \centering
    \inputdraft{
    
    \includegraphics{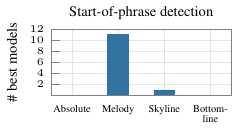}
    \hfill
    \includegraphics{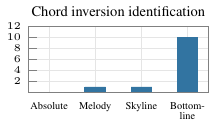}
    \includegraphics{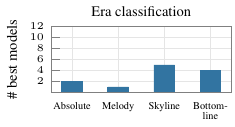}
    }{
        \input{figs/pgf_best_ref.tex}
    }
    \caption{
        Count of best intervalization references when comparing intervalized models with various references and the absolute.
        For each task, we consider two pre-trained and two end-to-end models trained on the tokenizations shown in Table~\ref{tab:tokenizations_description}. 
        This results in 12 comparisons by task, where each comparison involves three intervalization references tested against an absolute tokenization.
    }
    \label{fig:results_best_ref}
\end{figure}

%% file: figs/pgf_best_ref.tex
\pgfplotsset{%
  every axis legend/.append style ={%
    anchor = north west,%
    at = {(0.02,0.97)}%
  },
  /pgf/number format/.cd,
  use comma,
  1000 sep = {\,},
  min exponent for 1000 sep = 4,
}

\pgfplotsset{
  discard if/.style 2 args={
    x filter/.code={
      \edef\tempa{\thisrow{#1}}
      \edef\tempb{#2}
      \ifx\tempa\tempb
      \def\pgfmathresult{inf}
      \fi
    }
  },
  discard if not/.style 2 args={
    x filter/.code={
      \edef\tempa{\thisrow{#1}}
      \edef\tempb{#2}
      \ifx\tempa\tempb
      \else
      \def\pgfmathresult{inf}
      \fi
    }
  }
}

\makeatletter \newcommand{\pgfplotsdrawaxis}{\pgfplots@draw@axis} \makeatother

\pgfplotsset{axis line on top/.style={
  axis line style=transparent,
  ticklabel style=transparent,
  tick style=transparent,
  axis on top=false,
  after end axis/.append code={
    \pgfplotsset{axis line style=opaque,
      ticklabel style=opaque,
      tick style=opaque,
      grid=none}
    \pgfplotsdrawaxis}
  }
}

\definecolor{tab-1}{HTML}{3274a1} %
\definecolor{tab-2}{HTML}{e1802c} %
\definecolor{tab-3}{HTML}{2ca02c} %
\definecolor{tab-4}{HTML}{d62728} %
\definecolor{tab-5}{HTML}{9467bd} %
\definecolor{tab-6}{HTML}{8c564b} %
\definecolor{tab-7}{HTML}{e377c2} %
\definecolor{tab-8}{HTML}{7f7f7f} %
\definecolor{tab-9}{HTML}{bcbd22} %
\definecolor{tab-10}{HTML}{17becf} %

\newcommand\tinier{\fontsize{5pt}{5pt}\selectfont}

\centering
\tikzsetnextfilename{pgf_phrase_segmentation_best}
\begin{tikzpicture}
  \begin{axis}[
      axis line on top,
      ybar,
      align=center,
      bar width=10pt,
      ylabel={\# best models},
      xlabel={},
      xtick=data,
      symbolic x coords={Absolute, Melody, Skyline, Bottom-line},
      xticklabels={Absolute, Melody, Skyline, Bottom-\\line},
      xticklabel style={align=center},
      ytick={2,4,6,8,10,12},
      ymin=0, ymax=12,
      enlarge x limits=0.15,
      tick pos=left,
      cycle list name=color list,
      width=.55\linewidth,height=2.7cm,
      title style={
            yshift=-.5em
        },
      title={\scriptsize Start-of-phrase detection},
      axis line style = {gray},
      x tick label style={
        rotate=0,
        anchor=north,
        font=\tinier,
        yshift=.3em
      },
      xtick style={draw=none},
      y tick label style={
        rotate=0,
        font=\tinier
      },
      y label style={
        at={(axis description cs:-.15,.5)},anchor=south,
        font=\scriptsize
      },
      x label style={
        font=\scriptsize
      },
      grid=both,
      grid style={line width=.1pt, draw=gray!20},
    ]

    \addplot+[
    tab-1,fill=tab-1,  
    ]
    table[
      col sep=comma,
      x=reference_txt,
      y=best,
    ] {data/phrase_segmentation_best.csv};

  \end{axis}
\end{tikzpicture}
\tikzsetnextfilename{pgf_chord_identification_best}
\begin{tikzpicture}
  \begin{axis}[
      axis line on top,
      ybar,
      align=center,
      bar width=10pt,
      ylabel={},
      xlabel={},
      xtick=data,
      symbolic x coords={Absolute, Melody, Skyline, Bottom-line},
      xticklabels={Absolute, Melody, Skyline, Bottom-\\line},
      xticklabel style={align=center},
      ytick={2,4,6,8,10,12},
      ymin=0, ymax=12,
      enlarge x limits=0.15,
      tick pos=left,
      cycle list name=color list,
      width=.55\linewidth,height=2.7cm,
      title style={
            yshift=-.5em
        },
      title={\scriptsize Chord inversion identification},
      axis line style = {gray},
      x tick label style={
        rotate=0,
        anchor=north,
        font=\tinier,
        yshift=.3em
      },
      xtick style={draw=none},
      y tick label style={
        rotate=0,
        font=\tinier
      },
      y label style={
        at={(axis description cs:-.8,.5)},anchor=south,
        font=\scriptsize
      },
      x label style={
        font=\scriptsize
      },
      grid=both,
      grid style={line width=.1pt, draw=gray!20},
    ]

    \addplot+[
    tab-1,fill=tab-1,  
    ]
    table[
      col sep=comma,
      x=reference_txt,
      y=best,
    ] {data/chord_inversion_best.csv};

  \end{axis}
\end{tikzpicture}
\tikzsetnextfilename{pgf_era_clf_best}
\begin{tikzpicture}
  \begin{axis}[
      axis line on top,
      ybar,
      align=center,
      bar width=10pt,
      ylabel={\# best models},
      xlabel={},
      xtick=data,
      symbolic x coords={Absolute, Melody, Skyline, Bottom-line},
      xticklabels={Absolute, Melody, Skyline, Bottom-\\line},
      xticklabel style={align=center},
      ytick={2,4,6,8,10,12},
      ymin=0, ymax=12,
      enlarge x limits=0.15,
      tick pos=left,
      cycle list name=color list,
      width=.55\linewidth,height=2.7cm,
      title style={
            yshift=-.5em
        },
      title={\scriptsize Era classification},
      axis line style = {gray},
      x tick label style={
        rotate=0,
        anchor=north,
        font=\tinier,
        yshift=.3em
      },
      xtick style={draw=none},
      y tick label style={
        rotate=0,
        font=\tinier
      },
      y label style={
        at={(axis description cs:-.15,.5)},anchor=south,
        font=\scriptsize
      },
      x label style={
        font=\scriptsize
      },
      grid=both,
      grid style={line width=.1pt, draw=gray!20},
    ]

    \addplot+[
    tab-1,fill=tab-1,  
    ]
    table[
      col sep=comma,
      x=reference_txt,
      y=best,
    ] {data/era_clf_best.csv};

  \end{axis}
\end{tikzpicture}

%% file: figs/fig_inversion_interpretability.tex
\input{figs/pgf_header.tex}
\begin{figure}[t]
  \centering
  \begin{subfigure}[c]{\linewidth}
    \centering
    \subcaption{Root position}
    \label{fig:freq_inversion0}
    \begin{subfigure}[t]{.15\linewidth}
      \centering
      \includegraphics[width=\linewidth,valign=t]{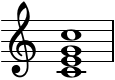}
      \vspace{-.1em}

      \resizebox{\linewidth}{!}{
        \color{red}
        $
        \begin{pmatrix}
          \text{Perf. oct.}\\
          \text{Perf. 5th}\\
          \text{Maj./min. 3rd}
        \end{pmatrix}
        $
      }
    \end{subfigure}
    \inputdraft{
      \raisebox{-.7\height}{\includegraphics{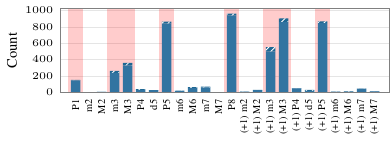}}
    }{
      \raisebox{-.7\height}{\input{figs/pgf_inversion0.tex}}
    }
  \end{subfigure}

  \begin{subfigure}[c]{\linewidth}
    \centering
    \subcaption{First inversion}
    \label{fig:freq_inversion1}
    \begin{subfigure}[t]{.15\linewidth}
      \centering
      \includegraphics[width=\linewidth,valign=t]{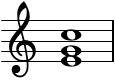}
      \vspace{-.1em}

      \resizebox{\linewidth}{!}{
        \color{red}
        $
        \begin{pmatrix}
          \text{Maj./min. 6th}\\
          \text{Maj./min. 3rd}
        \end{pmatrix}
        $
      }
    \end{subfigure}
    \inputdraft{
      \raisebox{-.7\height}{\includegraphics{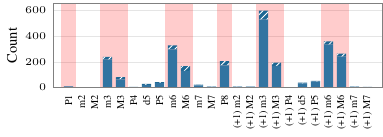}}
    }{
      \raisebox{-.7\height}{\input{figs/pgf_inversion1.tex}}
    }
  \end{subfigure}

  \begin{subfigure}[c]{\linewidth}
    \centering
    \subcaption{Second inversion}
    \label{fig:freq_inversion2}
    \begin{subfigure}[t]{.15\linewidth}
      \centering
      \includegraphics[width=\linewidth,valign=t]{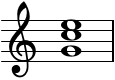}
      \vspace{-.1em}

      \resizebox{\linewidth}{!}{
        \color{red}
        $
        \begin{pmatrix}
          \text{Maj./min. 6th}\\
          \text{Perf. 4th}
        \end{pmatrix}
        $
      }
    \end{subfigure}
    \inputdraft{
      \raisebox{-.7\height}{\includegraphics{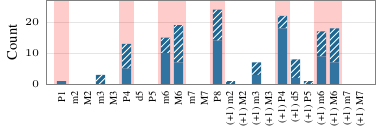}}
    }{
      \raisebox{-.7\height}{\input{figs/pgf_inversion2.tex}}
    }
  \end{subfigure}

  \begin{subfigure}[c]{\linewidth}
    \centering
    \subcaption{Third inversion}
    \label{fig:freq_inversion3}
    \begin{subfigure}[t]{.15\linewidth}
      \centering
      \includegraphics[width=\linewidth,valign=t]{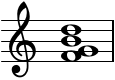}
      \vspace{-.1em}

      \resizebox{.8\linewidth}{!}{
        \color{red}
        $
        \begin{pmatrix}
          \text{Major 6th}\\
          \text{Dim. 5th}\\
          \text{Major 2nd}\\
        \end{pmatrix}
        $
      }
    \end{subfigure}
    \inputdraft{
      \raisebox{-.7\height}{\includegraphics{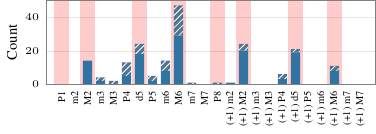}}
    }{
      \raisebox{-.7\height}{\input{figs/pgf_inversion3.tex}}
    }
  \end{subfigure}

  \caption{
    Histograms of vertical pitch interval tokens
    predicted as root position, first, second or third inversion.
    The tokenizer is a REMI intervalized tokenizer with the reference being the bottom-line encoded as absolute pitches and non-reference events encoded using vertical pitch intervals.
    The hatched part of a bar represents the proportion of false positives.
    Red highlights indicate the intervals that occur in each chord inversion.
    The notation (+1) indicates an additional octave.
  }
  \label{fig:inversion_interpretability}
\end{figure}

%% file: figs/pgf_header.tex
\pgfplotsset{%
  every axis legend/.append style ={%
    anchor = north west,%
    at = {(0.02,0.97)}%
  },
  /pgf/number format/.cd,
  use comma,
  1000 sep = {\,},
  min exponent for 1000 sep = 4,
}
\pgfplotsset{
  discard if/.style 2 args={
    x filter/.code={
      \edef\tempa{\thisrow{#1}}
      \edef\tempb{#2}
      \ifx\tempa\tempb
      \def\pgfmathresult{inf}
      \fi
    }
  },
  discard if not/.style 2 args={
    x filter/.code={
      \edef\tempa{\thisrow{#1}}
      \edef\tempb{#2}
      \ifx\tempa\tempb
      \else
      \def\pgfmathresult{inf}
      \fi
    }
  }
}
\makeatletter \newcommand{\pgfplotsdrawaxis}{\pgfplots@draw@axis} \makeatother
\pgfplotsset{axis line on top/.style={
    axis line style=transparent,
    ticklabel style=transparent,
    tick style=transparent,
    axis on top=false,
    after end axis/.append code={
      \pgfplotsset{axis line style=opaque,
        ticklabel style=opaque,
        tick style=opaque,
      grid=none}
    \pgfplotsdrawaxis}
  }
}
\definecolor{tab-1}{HTML}{3274a1} %
\definecolor{tab-2}{HTML}{e1802c} %
\definecolor{tab-3}{HTML}{2ca02c} %
\definecolor{tab-4}{HTML}{d62728} %
\definecolor{tab-5}{HTML}{9467bd} %
\definecolor{tab-6}{HTML}{8c564b} %
\definecolor{tab-7}{HTML}{e377c2} %
\definecolor{tab-8}{HTML}{7f7f7f} %
\definecolor{tab-9}{HTML}{bcbd22} %
\definecolor{tab-10}{HTML}{17becf} %
\definecolor{tab-1dark}{HTML}{315b78} %
\definecolor{tab-1light}{HTML}{527491} %
\newcommand\tinier{\fontsize{5pt}{5pt}\selectfont}
\newcommand\tinierer{\fontsize{4pt}{4pt}\selectfont}

%% file: figs/pgf_inversion0.tex
\pgfplotstableread[
  col sep=comma,
]{data/inversion0.csv}\inversionzerodata
\tikzsetnextfilename{pgf_inversion0}
\begin{tikzpicture}
  \begin{axis}[
      ybar stacked,
      axis line on top,
      bar width=4.5pt,
      ymin=0, ymax=1020,
      enlarge x limits=0.05,
      width=.85\linewidth,height=3cm,
      xticklabels from table={\inversionzerodata}{label},
      xticklabel style={
        rotate=90,
        anchor=east,
        font=\tinier,
      },
      xtick=data,
      ytick style={draw=none},
      xtick style={draw=none},
      ylabel={Count},
      y label style={
        font=\scriptsize,
        yshift=-.3em
      },
      y tick label style={
        rotate=0,
        font=\tinierer,
      },
      ymajorgrids=true,
      grid style={line width=.1pt, draw=gray!20},
      axis line style = {gray},
    ]
    \addplot [
      draw=none, fill=tab-1
    ] table[
      col sep=comma,
      meta=label,
      y=true_pos,
      x expr=\coordindex
    ] {\inversionzerodata};

    \addplot [
        draw=none, fill=tab-1light,
        postaction={
          pattern = north east lines, pattern color=white,
        }
    ] table[
      col sep=comma,
      meta=label,
      y=false_pos,
      x expr=\coordindex
    ] {\inversionzerodata};

    \fill [red, opacity=0.2] (-0.55, 0) rectangle (.55, 1200);
    \fill [red, opacity=0.2] (2.45, 0) rectangle (4.55, 1200);
    \fill [red, opacity=0.2] (6.45, 0) rectangle (7.55, 1200);
    \fill [red, opacity=0.2] (11.45, 0) rectangle (12.55, 1200);
    \fill [red, opacity=0.2] (14.45, 0) rectangle (16.55, 1200);
    \fill [red, opacity=0.2] (18.45, 0) rectangle (19.55, 1200);

  \end{axis}
\end{tikzpicture}

%% file: figs/pgf_inversion1.tex
\pgfplotstableread[
  col sep=comma,
]{data/inversion1.csv}\inversiononedata
\tikzsetnextfilename{pgf_inversion1}
\begin{tikzpicture}
  \begin{axis}[
      ybar stacked,
      axis line on top,
      bar width=4.5pt,
      ymin=0, ymax=650,
      enlarge x limits=0.05,
      width=.85\linewidth,height=3cm,
      xticklabels from table={\inversiononedata}{label},
      xticklabel style={
        rotate=90,
        anchor=east,
        font=\tinier,
      },
      xtick=data,
      ytick style={draw=none},
      xtick style={draw=none},
      ylabel={Count},
      y label style={
        font=\scriptsize,
        yshift=-.3em
      },
      y tick label style={
        rotate=0,
        font=\tinierer,
      },
      ymajorgrids=true,
      grid style={line width=.1pt, draw=gray!20},
      axis line style = {gray},
    ]
    \addplot [
      draw=none, fill=tab-1
    ] table[
      col sep=comma,
      meta=label,
      y=true_pos,
      x expr=\coordindex
    ] {\inversiononedata};

    \addplot [
        draw=none, fill=tab-1light,
        postaction={
          pattern = north east lines, pattern color=white,
        }
    ] table[
      col sep=comma,
      meta=label,
      y=false_pos,
      x expr=\coordindex
    ] {\inversiononedata};

    \fill [red, opacity=0.2] (-0.55, 0) rectangle (.55, 700);
    \fill [red, opacity=0.2] (2.45, 0) rectangle (4.55, 700);
    \fill [red, opacity=0.2] (7.45, 0) rectangle (9.55, 700);
    \fill [red, opacity=0.2] (11.45, 0) rectangle (12.55, 700);
    \fill [red, opacity=0.2] (14.45, 0) rectangle (16.55, 700);
    \fill [red, opacity=0.2] (19.45, 0) rectangle (21.55, 700);

  \end{axis}
\end{tikzpicture}

%% file: figs/pgf_inversion2.tex
\pgfplotstableread[
  col sep=comma,
]{data/inversion2.csv}\inversiontwodata
\tikzsetnextfilename{pgf_inversion2}
\begin{tikzpicture}
  \begin{axis}[
      ybar stacked,
      axis line on top,
      bar width=4.5pt,
      ymin=0, ymax=27,
      enlarge x limits=0.05,
      width=.85\linewidth,height=3cm,
      xticklabels from table={\inversiontwodata}{label},
      xticklabel style={
        rotate=90,
        anchor=east,
        font=\tinier,
      },
      xtick=data,
      ytick style={draw=none},
      xtick style={draw=none},
      ylabel={Count},
      y label style={
        font=\scriptsize,
        yshift=-.3em
      },
      y tick label style={
        rotate=0,
        font=\tinierer,
      },
      ymajorgrids=true,
      grid style={line width=.1pt, draw=gray!20},
      axis line style = {gray},
    ]
    \addplot [
      draw=none, fill=tab-1,
    ] table[
      col sep=comma,
      meta=label,
      y=true_pos,
      x expr=\coordindex
    ] {\inversiontwodata};

    \addplot [
        draw=none, fill=tab-1light,
        postaction={
          pattern = north east lines, pattern color=white,
        }
    ] table[
      col sep=comma,
      meta=label,
      y=false_pos,
      x expr=\coordindex
    ] {\inversiontwodata};

    \fill [red, opacity=0.2] (-0.55, 0) rectangle (.55, 30);
    \fill [red, opacity=0.2] (4.45, 0) rectangle (5.55, 30);
    \fill [red, opacity=0.2] (7.45, 0) rectangle (9.55, 30);
    \fill [red, opacity=0.2] (11.45, 0) rectangle (12.55, 30);
    \fill [red, opacity=0.2] (16.45, 0) rectangle (17.55, 30);
    \fill [red, opacity=0.2] (19.45, 0) rectangle (21.55, 30);

  \end{axis}
\end{tikzpicture}

%% file: figs/pgf_inversion3.tex
\pgfplotstableread[
  col sep=comma,
]{data/inversion3.csv}\inversionthreedata
\tikzsetnextfilename{pgf_inversion3}
\begin{tikzpicture}
  \begin{axis}[
      ybar stacked,
      axis line on top,
      bar width=4.5pt,
      ymin=0, ymax=50,
      enlarge x limits=0.05,
      width=.85\linewidth,height=3cm,
      xticklabels from table={\inversionthreedata}{label},
      xticklabel style={
        rotate=90,
        anchor=east,
        font=\tinier,
      },
      xtick=data,
      ytick style={draw=none},
      xtick style={draw=none},
      ylabel={Count},
      y label style={
        font=\scriptsize,
        yshift=-.3em
      },
      y tick label style={
        rotate=0,
        font=\tinierer,
      },
      ymajorgrids=true,
      grid style={line width=.1pt, draw=gray!20},
      axis line style = {gray},
    ]
    \addplot [
       draw=none, fill=tab-1,
    ] table[
      col sep=comma,
      meta=label,
      y=true_pos,
      x expr=\coordindex
    ] {\inversionthreedata};

    \addplot [
        draw=none, fill=tab-1light,
        postaction={
          pattern = north east lines, pattern color=white,
        }
    ] table[
      col sep=comma,
      meta=label,
      y=false_pos,
      x expr=\coordindex
    ] {\inversionthreedata};

    \fill [red, opacity=0.2] (-0.55, 0) rectangle (.55, 60);
    \fill [red, opacity=0.2] (1.45, 0) rectangle (2.55, 60);
    \fill [red, opacity=0.2] (5.45, 0) rectangle (6.55, 60);
    \fill [red, opacity=0.2] (8.45, 0) rectangle (9.55, 60);
    \fill [red, opacity=0.2] (11.45, 0) rectangle (12.55, 60);
    \fill [red, opacity=0.2] (13.45, 0) rectangle (14.55, 60);
    \fill [red, opacity=0.2] (17.45, 0) rectangle (18.55, 60);
    \fill [red, opacity=0.2] (20.45, 0) rectangle (21.55, 60);

  \end{axis}
\end{tikzpicture}

%% file: figs/fig_intervalization_future.tex
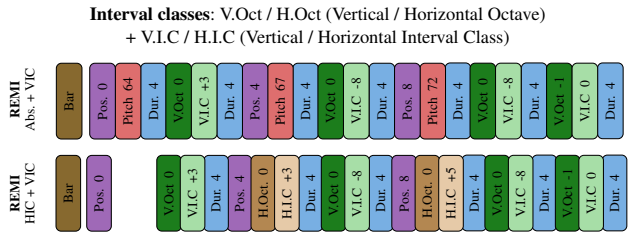
\begin{figure}[t]
    \centering
    {\scriptsize \textbf{Interval classes}: V.Oct / H.Oct (Vertical / Horizontal Octave)\\[-.3em]
    + V.I.C / H.I.C (Vertical / Horizontal Interval Class)}

    \vspace{.5em}

    \resizebox{\linewidth}{!}{
        \input{figs/tikz_intervalization_strategies_interval_class.tex}
    }

    \caption{
        Examples of intervalized tokenizations based on interval classes instead of pitch intervals.
        (Abs.: \textit{Absolute pitch encoding})
    }
    \label{fig:example_intervalization_future}
\end{figure}

%% file: figs/tikz_intervalization_strategies_interval_class.tex
\tikzset{every picture/.style={line width=0.75pt}} %

\begin{tikzpicture}[x=0.75pt,y=0.75pt,yscale=-1,xscale=1]

\draw  [fill={rgb, 255:red, 120; green, 175; blue, 227 }  ,fill opacity=1 ]  (132.81,11.07) .. controls (132.81,8.31) and (135.05,6.07) .. (137.81,6.07) -- (152.81,6.07) .. controls (155.57,6.07) and (157.81,8.31) .. (157.81,11.07) -- (157.81,75.07) .. controls (157.81,77.83) and (155.57,80.07) .. (152.81,80.07) -- (137.81,80.07) .. controls (135.05,80.07) and (132.81,77.83) .. (132.81,75.07) -- cycle  ;
\draw (145.31,43.07) node  [rotate=-270] [align=left] {\begin{minipage}[lt]{47.7pt}\setlength\topsep{0pt}
\begin{center}
Dur. \ 4
\end{center}

\end{minipage}};
\draw  [fill={rgb, 255:red, 120; green, 175; blue, 227 }  ,fill opacity=1 ]  (281.31,11.07) .. controls (281.31,8.31) and (283.55,6.07) .. (286.31,6.07) -- (301.31,6.07) .. controls (304.08,6.07) and (306.31,8.31) .. (306.31,11.07) -- (306.31,75.07) .. controls (306.31,77.83) and (304.08,80.07) .. (301.31,80.07) -- (286.31,80.07) .. controls (283.55,80.07) and (281.31,77.83) .. (281.31,75.07) -- cycle  ;
\draw (293.81,43.07) node  [rotate=-270] [align=left] {\begin{minipage}[lt]{47.7pt}\setlength\topsep{0pt}
\begin{center}
Dur. \ 4
\end{center}

\end{minipage}};
\draw  [fill={rgb, 255:red, 120; green, 175; blue, 227 }  ,fill opacity=1 ]  (355.82,11.07) .. controls (355.82,8.31) and (358.05,6.07) .. (360.82,6.07) -- (375.82,6.07) .. controls (378.58,6.07) and (380.82,8.31) .. (380.82,11.07) -- (380.82,75.07) .. controls (380.82,77.83) and (378.58,80.07) .. (375.82,80.07) -- (360.82,80.07) .. controls (358.05,80.07) and (355.82,77.83) .. (355.82,75.07) -- cycle  ;
\draw (368.32,43.07) node  [rotate=-270] [align=left] {\begin{minipage}[lt]{47.7pt}\setlength\topsep{0pt}
\begin{center}
Dur. \ 4
\end{center}

\end{minipage}};
\draw  [fill={rgb, 255:red, 120; green, 175; blue, 227 }  ,fill opacity=1 ]  (429.95,10.78) .. controls (429.95,8.02) and (432.19,5.78) .. (434.95,5.78) -- (449.95,5.78) .. controls (452.71,5.78) and (454.95,8.02) .. (454.95,10.78) -- (454.95,74.78) .. controls (454.95,77.54) and (452.71,79.78) .. (449.95,79.78) -- (434.95,79.78) .. controls (432.19,79.78) and (429.95,77.54) .. (429.95,74.78) -- cycle  ;
\draw (442.45,42.78) node  [rotate=-270] [align=left] {\begin{minipage}[lt]{47.7pt}\setlength\topsep{0pt}
\begin{center}
Dur. \ 4
\end{center}

\end{minipage}};
\draw  [fill={rgb, 255:red, 222; green, 110; blue, 110 }  ,fill opacity=1 ]  (108.06,11.07) .. controls (108.06,8.31) and (110.3,6.07) .. (113.06,6.07) -- (128.06,6.07) .. controls (130.82,6.07) and (133.06,8.31) .. (133.06,11.07) -- (133.06,75.07) .. controls (133.06,77.83) and (130.82,80.07) .. (128.06,80.07) -- (113.06,80.07) .. controls (110.3,80.07) and (108.06,77.83) .. (108.06,75.07) -- cycle  ;
\draw (120.56,43.07) node  [rotate=-270] [align=left] {\begin{minipage}[lt]{47.7pt}\setlength\topsep{0pt}
\begin{center}
Pitch \ 64
\end{center}

\end{minipage}};
\draw  [fill={rgb, 255:red, 222; green, 110; blue, 110 }  ,fill opacity=1 ]  (256.68,11.07) .. controls (256.68,8.31) and (258.91,6.07) .. (261.68,6.07) -- (276.68,6.07) .. controls (279.44,6.07) and (281.68,8.31) .. (281.68,11.07) -- (281.68,75.07) .. controls (281.68,77.83) and (279.44,80.07) .. (276.68,80.07) -- (261.68,80.07) .. controls (258.91,80.07) and (256.68,77.83) .. (256.68,75.07) -- cycle  ;
\draw (269.18,43.07) node  [rotate=-270] [align=left] {\begin{minipage}[lt]{47.7pt}\setlength\topsep{0pt}
\begin{center}
Pitch \ 67
\end{center}

\end{minipage}};
\draw  [fill={rgb, 255:red, 222; green, 110; blue, 110 }  ,fill opacity=1 ]  (405.37,11.07) .. controls (405.37,8.31) and (407.61,6.07) .. (410.37,6.07) -- (425.37,6.07) .. controls (428.13,6.07) and (430.37,8.31) .. (430.37,11.07) -- (430.37,75.07) .. controls (430.37,77.83) and (428.13,80.07) .. (425.37,80.07) -- (410.37,80.07) .. controls (407.61,80.07) and (405.37,77.83) .. (405.37,75.07) -- cycle  ;
\draw (417.87,43.07) node  [rotate=-270] [align=left] {\begin{minipage}[lt]{47.7pt}\setlength\topsep{0pt}
\begin{center}
Pitch \ 72
\end{center}

\end{minipage}};
\draw  [fill={rgb, 255:red, 159; green, 105; blue, 183 }  ,fill opacity=1 ]  (83.14,11.07) .. controls (83.14,8.31) and (85.38,6.07) .. (88.14,6.07) -- (103.14,6.07) .. controls (105.9,6.07) and (108.14,8.31) .. (108.14,11.07) -- (108.14,75.07) .. controls (108.14,77.83) and (105.9,80.07) .. (103.14,80.07) -- (88.14,80.07) .. controls (85.38,80.07) and (83.14,77.83) .. (83.14,75.07) -- cycle  ;
\draw (95.64,43.07) node  [rotate=-270] [align=left] {\begin{minipage}[lt]{47.7pt}\setlength\topsep{0pt}
\begin{center}
Pos. \ 0
\end{center}

\end{minipage}};
\draw  [fill={rgb, 255:red, 135; green, 105; blue, 47 }  ,fill opacity=1 ]  (50.56,11.07) .. controls (50.56,8.31) and (52.8,6.07) .. (55.56,6.07) -- (70.56,6.07) .. controls (73.33,6.07) and (75.56,8.31) .. (75.56,11.07) -- (75.56,75.07) .. controls (75.56,77.83) and (73.33,80.07) .. (70.56,80.07) -- (55.56,80.07) .. controls (52.8,80.07) and (50.56,77.83) .. (50.56,75.07) -- cycle  ;
\draw (63.06,43.07) node  [rotate=-270] [align=left] {\begin{minipage}[lt]{47.7pt}\setlength\topsep{0pt}
\begin{center}
Bar
\end{center}

\end{minipage}};
\draw  [fill={rgb, 255:red, 159; green, 105; blue, 183 }  ,fill opacity=1 ]  (231.99,11.07) .. controls (231.99,8.31) and (234.23,6.07) .. (236.99,6.07) -- (251.99,6.07) .. controls (254.75,6.07) and (256.99,8.31) .. (256.99,11.07) -- (256.99,75.07) .. controls (256.99,77.83) and (254.75,80.07) .. (251.99,80.07) -- (236.99,80.07) .. controls (234.23,80.07) and (231.99,77.83) .. (231.99,75.07) -- cycle  ;
\draw (244.49,43.07) node  [rotate=-270] [align=left] {\begin{minipage}[lt]{47.7pt}\setlength\topsep{0pt}
\begin{center}
Pos. \ 4
\end{center}

\end{minipage}};
\draw  [fill={rgb, 255:red, 159; green, 105; blue, 183 }  ,fill opacity=1 ]  (380.61,11.07) .. controls (380.61,8.31) and (382.84,6.07) .. (385.61,6.07) -- (400.61,6.07) .. controls (403.37,6.07) and (405.61,8.31) .. (405.61,11.07) -- (405.61,75.07) .. controls (405.61,77.83) and (403.37,80.07) .. (400.61,80.07) -- (385.61,80.07) .. controls (382.84,80.07) and (380.61,77.83) .. (380.61,75.07) -- cycle  ;
\draw (393.11,43.07) node  [rotate=-270] [align=left] {\begin{minipage}[lt]{47.7pt}\setlength\topsep{0pt}
\begin{center}
Pos. \ 8
\end{center}

\end{minipage}};
\draw (4,77.3) node [anchor=north west][inner sep=0.75pt]  [rotate=-270] [align=left] {\begin{minipage}[lt]{51.76pt}\setlength\topsep{0pt}
\begin{center}
\textbf{REMI}\\Abs. + VIC
\end{center}

\end{minipage}};
\draw  [fill={rgb, 255:red, 120; green, 175; blue, 227 }  ,fill opacity=1 ]  (207.2,11.07) .. controls (207.2,8.31) and (209.43,6.07) .. (212.2,6.07) -- (227.2,6.07) .. controls (229.96,6.07) and (232.2,8.31) .. (232.2,11.07) -- (232.2,75.07) .. controls (232.2,77.83) and (229.96,80.07) .. (227.2,80.07) -- (212.2,80.07) .. controls (209.43,80.07) and (207.2,77.83) .. (207.2,75.07) -- cycle  ;
\draw (219.7,43.07) node  [rotate=-270] [align=left] {\begin{minipage}[lt]{47.7pt}\setlength\topsep{0pt}
\begin{center}
Dur. \ 4
\end{center}

\end{minipage}};
\draw  [fill={rgb, 255:red, 120; green, 175; blue, 227 }  ,fill opacity=1 ]  (504.33,10.78) .. controls (504.33,8.02) and (506.57,5.78) .. (509.33,5.78) -- (524.33,5.78) .. controls (527.09,5.78) and (529.33,8.02) .. (529.33,10.78) -- (529.33,74.78) .. controls (529.33,77.54) and (527.09,79.78) .. (524.33,79.78) -- (509.33,79.78) .. controls (506.57,79.78) and (504.33,77.54) .. (504.33,74.78) -- cycle  ;
\draw (516.83,42.78) node  [rotate=-270] [align=left] {\begin{minipage}[lt]{47.7pt}\setlength\topsep{0pt}
\begin{center}
Dur. \ 4
\end{center}

\end{minipage}};
\draw  [fill={rgb, 255:red, 120; green, 175; blue, 227 }  ,fill opacity=1 ]  (578.72,10.78) .. controls (578.72,8.02) and (580.95,5.78) .. (583.72,5.78) -- (598.72,5.78) .. controls (601.48,5.78) and (603.72,8.02) .. (603.72,10.78) -- (603.72,74.78) .. controls (603.72,77.54) and (601.48,79.78) .. (598.72,79.78) -- (583.72,79.78) .. controls (580.95,79.78) and (578.72,77.54) .. (578.72,74.78) -- cycle  ;
\draw (591.22,42.78) node  [rotate=-270] [align=left] {\begin{minipage}[lt]{47.7pt}\setlength\topsep{0pt}
\begin{center}
Dur. \ 4
\end{center}

\end{minipage}};
\draw  [fill={rgb, 255:red, 28; green, 125; blue, 28 }  ,fill opacity=1 ]  (157.43,11) .. controls (157.43,8.24) and (159.66,6) .. (162.43,6) -- (177.43,6) .. controls (180.19,6) and (182.43,8.24) .. (182.43,11) -- (182.43,75) .. controls (182.43,77.76) and (180.19,80) .. (177.43,80) -- (162.43,80) .. controls (159.66,80) and (157.43,77.76) .. (157.43,75) -- cycle  ;
\draw (169.93,43) node  [rotate=-270] [align=left] {\begin{minipage}[lt]{47.78pt}\setlength\topsep{0pt}
\begin{center}
V.Oct \ 0
\end{center}

\end{minipage}};
\draw  [fill={rgb, 255:red, 167; green, 222; blue, 167 }  ,fill opacity=1 ]  (182.32,11) .. controls (182.32,8.24) and (184.56,6) .. (187.32,6) -- (202.32,6) .. controls (205.09,6) and (207.32,8.24) .. (207.32,11) -- (207.32,75) .. controls (207.32,77.76) and (205.09,80) .. (202.32,80) -- (187.32,80) .. controls (184.56,80) and (182.32,77.76) .. (182.32,75) -- cycle  ;
\draw (194.82,43) node  [rotate=-270] [align=left] {\begin{minipage}[lt]{47.78pt}\setlength\topsep{0pt}
\begin{center}
V.I.C \ +3
\end{center}

\end{minipage}};
\draw  [fill={rgb, 255:red, 28; green, 125; blue, 28 }  ,fill opacity=1 ]  (305.86,11) .. controls (305.86,8.24) and (308.1,6) .. (310.86,6) -- (325.86,6) .. controls (328.62,6) and (330.86,8.24) .. (330.86,11) -- (330.86,75) .. controls (330.86,77.76) and (328.62,80) .. (325.86,80) -- (310.86,80) .. controls (308.1,80) and (305.86,77.76) .. (305.86,75) -- cycle  ;
\draw (318.36,43) node  [rotate=-270] [align=left] {\begin{minipage}[lt]{47.78pt}\setlength\topsep{0pt}
\begin{center}
V.Oct \ 0
\end{center}

\end{minipage}};
\draw  [fill={rgb, 255:red, 167; green, 222; blue, 167 }  ,fill opacity=1 ]  (330.76,11) .. controls (330.76,8.24) and (332.99,6) .. (335.76,6) -- (350.76,6) .. controls (353.52,6) and (355.76,8.24) .. (355.76,11) -- (355.76,75) .. controls (355.76,77.76) and (353.52,80) .. (350.76,80) -- (335.76,80) .. controls (332.99,80) and (330.76,77.76) .. (330.76,75) -- cycle  ;
\draw (343.26,43) node  [rotate=-270] [align=left] {\begin{minipage}[lt]{47.78pt}\setlength\topsep{0pt}
\begin{center}
V.I.C \ -8
\end{center}

\end{minipage}};
\draw  [fill={rgb, 255:red, 28; green, 125; blue, 28 }  ,fill opacity=1 ]  (454.29,11) .. controls (454.29,8.24) and (456.53,6) .. (459.29,6) -- (474.29,6) .. controls (477.05,6) and (479.29,8.24) .. (479.29,11) -- (479.29,75) .. controls (479.29,77.76) and (477.05,80) .. (474.29,80) -- (459.29,80) .. controls (456.53,80) and (454.29,77.76) .. (454.29,75) -- cycle  ;
\draw (466.79,43) node  [rotate=-270] [align=left] {\begin{minipage}[lt]{47.78pt}\setlength\topsep{0pt}
\begin{center}
V.Oct \ 0
\end{center}

\end{minipage}};
\draw  [fill={rgb, 255:red, 167; green, 222; blue, 167 }  ,fill opacity=1 ]  (479.19,11) .. controls (479.19,8.24) and (481.43,6) .. (484.19,6) -- (499.19,6) .. controls (501.95,6) and (504.19,8.24) .. (504.19,11) -- (504.19,75) .. controls (504.19,77.76) and (501.95,80) .. (499.19,80) -- (484.19,80) .. controls (481.43,80) and (479.19,77.76) .. (479.19,75) -- cycle  ;
\draw (491.69,43) node  [rotate=-270] [align=left] {\begin{minipage}[lt]{47.78pt}\setlength\topsep{0pt}
\begin{center}
V.I.C \ -8
\end{center}

\end{minipage}};
\draw  [fill={rgb, 255:red, 28; green, 125; blue, 28 }  ,fill opacity=1 ]  (528.98,11) .. controls (528.98,8.24) and (531.22,6) .. (533.98,6) -- (548.98,6) .. controls (551.75,6) and (553.98,8.24) .. (553.98,11) -- (553.98,75) .. controls (553.98,77.76) and (551.75,80) .. (548.98,80) -- (533.98,80) .. controls (531.22,80) and (528.98,77.76) .. (528.98,75) -- cycle  ;
\draw (541.48,43) node  [rotate=-270] [align=left] {\begin{minipage}[lt]{47.78pt}\setlength\topsep{0pt}
\begin{center}
V.Oct \ -1
\end{center}

\end{minipage}};
\draw  [fill={rgb, 255:red, 167; green, 222; blue, 167 }  ,fill opacity=1 ]  (553.88,11) .. controls (553.88,8.24) and (556.12,6) .. (558.88,6) -- (573.88,6) .. controls (576.64,6) and (578.88,8.24) .. (578.88,11) -- (578.88,75) .. controls (578.88,77.76) and (576.64,80) .. (573.88,80) -- (558.88,80) .. controls (556.12,80) and (553.88,77.76) .. (553.88,75) -- cycle  ;
\draw (566.38,43) node  [rotate=-270] [align=left] {\begin{minipage}[lt]{47.78pt}\setlength\topsep{0pt}
\begin{center}
V.I.C \ 0
\end{center}

\end{minipage}};
\draw  [fill={rgb, 255:red, 120; green, 175; blue, 227 }  ,fill opacity=1 ]  (286.57,101.07) .. controls (286.57,98.31) and (288.8,96.07) .. (291.57,96.07) -- (305.57,96.07) .. controls (308.33,96.07) and (310.57,98.31) .. (310.57,101.07) -- (310.57,165.07) .. controls (310.57,167.83) and (308.33,170.07) .. (305.57,170.07) -- (291.57,170.07) .. controls (288.8,170.07) and (286.57,167.83) .. (286.57,165.07) -- cycle  ;
\draw (298.57,133.07) node  [rotate=-270] [align=left] {\begin{minipage}[lt]{47.7pt}\setlength\topsep{0pt}
\begin{center}
Dur. \ 4
\end{center}

\end{minipage}};
\draw  [fill={rgb, 255:red, 120; green, 175; blue, 227 }  ,fill opacity=1 ]  (355.14,101.07) .. controls (355.14,98.31) and (357.38,96.07) .. (360.14,96.07) -- (374.14,96.07) .. controls (376.91,96.07) and (379.14,98.31) .. (379.14,101.07) -- (379.14,165.07) .. controls (379.14,167.83) and (376.91,170.07) .. (374.14,170.07) -- (360.14,170.07) .. controls (357.38,170.07) and (355.14,167.83) .. (355.14,165.07) -- cycle  ;
\draw (367.14,133.07) node  [rotate=-270] [align=left] {\begin{minipage}[lt]{47.7pt}\setlength\topsep{0pt}
\begin{center}
Dur. \ 4
\end{center}

\end{minipage}};
\draw  [fill={rgb, 255:red, 120; green, 175; blue, 227 }  ,fill opacity=1 ]  (446.39,100.78) .. controls (446.39,98.02) and (448.63,95.78) .. (451.39,95.78) -- (465.39,95.78) .. controls (468.16,95.78) and (470.39,98.02) .. (470.39,100.78) -- (470.39,164.78) .. controls (470.39,167.54) and (468.16,169.78) .. (465.39,169.78) -- (451.39,169.78) .. controls (448.63,169.78) and (446.39,167.54) .. (446.39,164.78) -- cycle  ;
\draw (458.39,132.78) node  [rotate=-270] [align=left] {\begin{minipage}[lt]{47.7pt}\setlength\topsep{0pt}
\begin{center}
Dur. \ 4
\end{center}

\end{minipage}};
\draw  [fill={rgb, 255:red, 159; green, 105; blue, 183 }  ,fill opacity=1 ]  (80.22,101.07) .. controls (80.22,98.31) and (82.46,96.07) .. (85.22,96.07) -- (99.22,96.07) .. controls (101.98,96.07) and (104.22,98.31) .. (104.22,101.07) -- (104.22,165.07) .. controls (104.22,167.83) and (101.98,170.07) .. (99.22,170.07) -- (85.22,170.07) .. controls (82.46,170.07) and (80.22,167.83) .. (80.22,165.07) -- cycle  ;
\draw (92.22,133.07) node  [rotate=-270] [align=left] {\begin{minipage}[lt]{47.7pt}\setlength\topsep{0pt}
\begin{center}
Pos. \ 0
\end{center}

\end{minipage}};
\draw  [fill={rgb, 255:red, 135; green, 105; blue, 47 }  ,fill opacity=1 ]  (50.23,101.07) .. controls (50.23,98.31) and (52.47,96.07) .. (55.23,96.07) -- (69.23,96.07) .. controls (71.99,96.07) and (74.23,98.31) .. (74.23,101.07) -- (74.23,165.07) .. controls (74.23,167.83) and (71.99,170.07) .. (69.23,170.07) -- (55.23,170.07) .. controls (52.47,170.07) and (50.23,167.83) .. (50.23,165.07) -- cycle  ;
\draw (62.23,133.07) node  [rotate=-270] [align=left] {\begin{minipage}[lt]{47.7pt}\setlength\topsep{0pt}
\begin{center}
Bar
\end{center}

\end{minipage}};
\draw  [fill={rgb, 255:red, 159; green, 105; blue, 183 }  ,fill opacity=1 ]  (217.23,101.07) .. controls (217.23,98.31) and (219.47,96.07) .. (222.23,96.07) -- (236.23,96.07) .. controls (238.99,96.07) and (241.23,98.31) .. (241.23,101.07) -- (241.23,165.07) .. controls (241.23,167.83) and (238.99,170.07) .. (236.23,170.07) -- (222.23,170.07) .. controls (219.47,170.07) and (217.23,167.83) .. (217.23,165.07) -- cycle  ;
\draw (229.23,133.07) node  [rotate=-270] [align=left] {\begin{minipage}[lt]{47.7pt}\setlength\topsep{0pt}
\begin{center}
Pos. \ 4
\end{center}

\end{minipage}};
\draw  [fill={rgb, 255:red, 159; green, 105; blue, 183 }  ,fill opacity=1 ]  (377.96,101.07) .. controls (377.96,98.31) and (380.2,96.07) .. (382.96,96.07) -- (396.96,96.07) .. controls (399.72,96.07) and (401.96,98.31) .. (401.96,101.07) -- (401.96,165.07) .. controls (401.96,167.83) and (399.72,170.07) .. (396.96,170.07) -- (382.96,170.07) .. controls (380.2,170.07) and (377.96,167.83) .. (377.96,165.07) -- cycle  ;
\draw (389.96,133.07) node  [rotate=-270] [align=left] {\begin{minipage}[lt]{47.7pt}\setlength\topsep{0pt}
\begin{center}
Pos. \ 8
\end{center}

\end{minipage}};
\draw (4,167.3) node [anchor=north west][inner sep=0.75pt]  [rotate=-270] [align=left] {\begin{minipage}[lt]{48.91pt}\setlength\topsep{0pt}
\begin{center}
\textbf{REMI}\\HIC + VIC
\end{center}

\end{minipage}};
\draw  [fill={rgb, 255:red, 120; green, 175; blue, 227 }  ,fill opacity=1 ]  (194.41,101.07) .. controls (194.41,98.31) and (196.65,96.07) .. (199.41,96.07) -- (213.41,96.07) .. controls (216.17,96.07) and (218.41,98.31) .. (218.41,101.07) -- (218.41,165.07) .. controls (218.41,167.83) and (216.17,170.07) .. (213.41,170.07) -- (199.41,170.07) .. controls (196.65,170.07) and (194.41,167.83) .. (194.41,165.07) -- cycle  ;
\draw (206.41,133.07) node  [rotate=-270] [align=left] {\begin{minipage}[lt]{47.7pt}\setlength\topsep{0pt}
\begin{center}
Dur. \ 4
\end{center}

\end{minipage}};
\draw  [fill={rgb, 255:red, 120; green, 175; blue, 227 }  ,fill opacity=1 ]  (514.86,100.78) .. controls (514.86,98.02) and (517.1,95.78) .. (519.86,95.78) -- (533.86,95.78) .. controls (536.63,95.78) and (538.86,98.02) .. (538.86,100.78) -- (538.86,164.78) .. controls (538.86,167.54) and (536.63,169.78) .. (533.86,169.78) -- (519.86,169.78) .. controls (517.1,169.78) and (514.86,167.54) .. (514.86,164.78) -- cycle  ;
\draw (526.86,132.78) node  [rotate=-270] [align=left] {\begin{minipage}[lt]{47.7pt}\setlength\topsep{0pt}
\begin{center}
Dur. \ 4
\end{center}

\end{minipage}};
\draw  [fill={rgb, 255:red, 120; green, 175; blue, 227 }  ,fill opacity=1 ]  (583.33,100.78) .. controls (583.33,98.02) and (585.57,95.78) .. (588.33,95.78) -- (602.33,95.78) .. controls (605.1,95.78) and (607.33,98.02) .. (607.33,100.78) -- (607.33,164.78) .. controls (607.33,167.54) and (605.1,169.78) .. (602.33,169.78) -- (588.33,169.78) .. controls (585.57,169.78) and (583.33,167.54) .. (583.33,164.78) -- cycle  ;
\draw (595.33,132.78) node  [rotate=-270] [align=left] {\begin{minipage}[lt]{47.7pt}\setlength\topsep{0pt}
\begin{center}
Dur. \ 4
\end{center}

\end{minipage}};
\draw  [fill={rgb, 255:red, 28; green, 125; blue, 28 }  ,fill opacity=1 ]  (148.6,101) .. controls (148.6,98.24) and (150.83,96) .. (153.6,96) -- (167.6,96) .. controls (170.36,96) and (172.6,98.24) .. (172.6,101) -- (172.6,165) .. controls (172.6,167.76) and (170.36,170) .. (167.6,170) -- (153.6,170) .. controls (150.83,170) and (148.6,167.76) .. (148.6,165) -- cycle  ;
\draw (160.6,133) node  [rotate=-270] [align=left] {\begin{minipage}[lt]{47.78pt}\setlength\topsep{0pt}
\begin{center}
V.Oct \ 0
\end{center}

\end{minipage}};
\draw  [fill={rgb, 255:red, 167; green, 222; blue, 167 }  ,fill opacity=1 ]  (171.51,101) .. controls (171.51,98.24) and (173.75,96) .. (176.51,96) -- (190.51,96) .. controls (193.28,96) and (195.51,98.24) .. (195.51,101) -- (195.51,165) .. controls (195.51,167.76) and (193.28,170) .. (190.51,170) -- (176.51,170) .. controls (173.75,170) and (171.51,167.76) .. (171.51,165) -- cycle  ;
\draw (183.51,133) node  [rotate=-270] [align=left] {\begin{minipage}[lt]{47.78pt}\setlength\topsep{0pt}
\begin{center}
V.I.C \ +3
\end{center}

\end{minipage}};
\draw  [fill={rgb, 255:red, 28; green, 125; blue, 28 }  ,fill opacity=1 ]  (309.16,101) .. controls (309.16,98.24) and (311.4,96) .. (314.16,96) -- (328.16,96) .. controls (330.92,96) and (333.16,98.24) .. (333.16,101) -- (333.16,165) .. controls (333.16,167.76) and (330.92,170) .. (328.16,170) -- (314.16,170) .. controls (311.4,170) and (309.16,167.76) .. (309.16,165) -- cycle  ;
\draw (321.16,133) node  [rotate=-270] [align=left] {\begin{minipage}[lt]{47.78pt}\setlength\topsep{0pt}
\begin{center}
V.Oct \ 0
\end{center}

\end{minipage}};
\draw  [fill={rgb, 255:red, 167; green, 222; blue, 167 }  ,fill opacity=1 ]  (332.08,101) .. controls (332.08,98.24) and (334.32,96) .. (337.08,96) -- (351.08,96) .. controls (353.84,96) and (356.08,98.24) .. (356.08,101) -- (356.08,165) .. controls (356.08,167.76) and (353.84,170) .. (351.08,170) -- (337.08,170) .. controls (334.32,170) and (332.08,167.76) .. (332.08,165) -- cycle  ;
\draw (344.08,133) node  [rotate=-270] [align=left] {\begin{minipage}[lt]{47.78pt}\setlength\topsep{0pt}
\begin{center}
V.I.C \ -8
\end{center}

\end{minipage}};
\draw  [fill={rgb, 255:red, 28; green, 125; blue, 28 }  ,fill opacity=1 ]  (468.8,101) .. controls (468.8,98.24) and (471.04,96) .. (473.8,96) -- (487.8,96) .. controls (490.56,96) and (492.8,98.24) .. (492.8,101) -- (492.8,165) .. controls (492.8,167.76) and (490.56,170) .. (487.8,170) -- (473.8,170) .. controls (471.04,170) and (468.8,167.76) .. (468.8,165) -- cycle  ;
\draw (480.8,133) node  [rotate=-270] [align=left] {\begin{minipage}[lt]{47.78pt}\setlength\topsep{0pt}
\begin{center}
V.Oct \ 0
\end{center}

\end{minipage}};
\draw  [fill={rgb, 255:red, 167; green, 222; blue, 167 }  ,fill opacity=1 ]  (491.72,101) .. controls (491.72,98.24) and (493.96,96) .. (496.72,96) -- (510.72,96) .. controls (513.48,96) and (515.72,98.24) .. (515.72,101) -- (515.72,165) .. controls (515.72,167.76) and (513.48,170) .. (510.72,170) -- (496.72,170) .. controls (493.96,170) and (491.72,167.76) .. (491.72,165) -- cycle  ;
\draw (503.72,133) node  [rotate=-270] [align=left] {\begin{minipage}[lt]{47.78pt}\setlength\topsep{0pt}
\begin{center}
V.I.C \ -8
\end{center}

\end{minipage}};
\draw  [fill={rgb, 255:red, 28; green, 125; blue, 28 }  ,fill opacity=1 ]  (537.56,101) .. controls (537.56,98.24) and (539.8,96) .. (542.56,96) -- (556.56,96) .. controls (559.32,96) and (561.56,98.24) .. (561.56,101) -- (561.56,165) .. controls (561.56,167.76) and (559.32,170) .. (556.56,170) -- (542.56,170) .. controls (539.8,170) and (537.56,167.76) .. (537.56,165) -- cycle  ;
\draw (549.56,133) node  [rotate=-270] [align=left] {\begin{minipage}[lt]{47.78pt}\setlength\topsep{0pt}
\begin{center}
V.Oct \ -1
\end{center}

\end{minipage}};
\draw  [fill={rgb, 255:red, 167; green, 222; blue, 167 }  ,fill opacity=1 ]  (560.48,101) .. controls (560.48,98.24) and (562.71,96) .. (565.48,96) -- (579.48,96) .. controls (582.24,96) and (584.48,98.24) .. (584.48,101) -- (584.48,165) .. controls (584.48,167.76) and (582.24,170) .. (579.48,170) -- (565.48,170) .. controls (562.71,170) and (560.48,167.76) .. (560.48,165) -- cycle  ;
\draw (572.48,133) node  [rotate=-270] [align=left] {\begin{minipage}[lt]{47.78pt}\setlength\topsep{0pt}
\begin{center}
V.I.C \ 0
\end{center}

\end{minipage}};
\draw  [fill={rgb, 255:red, 179; green, 135; blue, 81 }  ,fill opacity=1 ]  (240.54,101.07) .. controls (240.54,98.31) and (242.78,96.07) .. (245.54,96.07) -- (259.54,96.07) .. controls (262.3,96.07) and (264.54,98.31) .. (264.54,101.07) -- (264.54,165.07) .. controls (264.54,167.83) and (262.3,170.07) .. (259.54,170.07) -- (245.54,170.07) .. controls (242.78,170.07) and (240.54,167.83) .. (240.54,165.07) -- cycle  ;
\draw (252.54,133.07) node  [rotate=-270] [align=left] {\begin{minipage}[lt]{47.7pt}\setlength\topsep{0pt}
\begin{center}
H.Oct. \ 0
\end{center}

\end{minipage}};
\draw  [fill={rgb, 255:red, 230; green, 202; blue, 168 }  ,fill opacity=1 ]  (263.55,101.07) .. controls (263.55,98.31) and (265.79,96.07) .. (268.55,96.07) -- (282.55,96.07) .. controls (285.31,96.07) and (287.55,98.31) .. (287.55,101.07) -- (287.55,165.07) .. controls (287.55,167.83) and (285.31,170.07) .. (282.55,170.07) -- (268.55,170.07) .. controls (265.79,170.07) and (263.55,167.83) .. (263.55,165.07) -- cycle  ;
\draw (275.55,133.07) node  [rotate=-270] [align=left] {\begin{minipage}[lt]{47.7pt}\setlength\topsep{0pt}
\begin{center}
H.I.C \ +3
\end{center}

\end{minipage}};
\draw  [fill={rgb, 255:red, 179; green, 135; blue, 81 }  ,fill opacity=1 ]  (400.71,101.07) .. controls (400.71,98.31) and (402.95,96.07) .. (405.71,96.07) -- (419.71,96.07) .. controls (422.47,96.07) and (424.71,98.31) .. (424.71,101.07) -- (424.71,165.07) .. controls (424.71,167.83) and (422.47,170.07) .. (419.71,170.07) -- (405.71,170.07) .. controls (402.95,170.07) and (400.71,167.83) .. (400.71,165.07) -- cycle  ;
\draw (412.71,133.07) node  [rotate=-270] [align=left] {\begin{minipage}[lt]{47.7pt}\setlength\topsep{0pt}
\begin{center}
H.Oct. \ 0
\end{center}

\end{minipage}};
\draw  [fill={rgb, 255:red, 230; green, 202; blue, 168 }  ,fill opacity=1 ]  (423.72,101.07) .. controls (423.72,98.31) and (425.96,96.07) .. (428.72,96.07) -- (442.72,96.07) .. controls (445.48,96.07) and (447.72,98.31) .. (447.72,101.07) -- (447.72,165.07) .. controls (447.72,167.83) and (445.48,170.07) .. (442.72,170.07) -- (428.72,170.07) .. controls (425.96,170.07) and (423.72,167.83) .. (423.72,165.07) -- cycle  ;
\draw (435.72,133.07) node  [rotate=-270] [align=left] {\begin{minipage}[lt]{47.7pt}\setlength\topsep{0pt}
\begin{center}
H.I.C \ +5
\end{center}

\end{minipage}};

\end{tikzpicture}